\pdfoutput=1
\documentclass[
	affil-it, 
	auth-lg, 
	british, 
	compression, 
	contents, 
	custom-packages={simpler-wick,tabularx}, 
	references={bigone}, 
]{lib/preprint}

\newcommand\whpl{{\widehat{+}}}
\newcommand\whmi{{\widehat{-}}}

\begin{document}
	\hypersetup{
		pdftitle = {Classification of Killing Horizons in D=11 Supergravity},
		pdfauthor = {Jan Gutowski,Chettha Saelim,Martin Wolf},
		pdfkeywords = {}
	}

	\date{\today}

	\email{j.gutowski@surrey.ac.uk,c.saelim@surrey.ac.uk,m.wolf@surrey.ac.uk}

	\preprint{DMUS--MP--26/02}

	\title{Classification of Killing Horizons in $D=11$ Supergravity}

	\author[a]{Jan~Gutowski\,\orcidlink{0000-0001-8807-3818}\,}
	\author[a]{Chettha~Saelim\,\orcidlink{0009-0006-8249-1334}\,}
	\author[a]{Martin~Wolf\,\orcidlink{0009-0002-8192-3124}\,}

	\affil[a]{School of Mathematics and Physics,\\ University of Surrey, Guildford GU2 7XH, United Kingdom}

	\abstract{We initiate the classification of supersymmetric degenerate Killing horizons, with closed spatial cross section, away from the near-horizon limit in $D=11$ supergravity. We prove that all such solutions fall into two distinct classes, depending on lightcone chirality with respect to a Gau{\ss}ian null coordinate system. For the first class of solutions, the negative lightcone chirality part of the Killing spinor is non-zero on the Killing horizon, and we prove that all such solutions are isometric to supersymmetric near-horizon geometries. In the second class, the negative lightcone chirality part of the Killing spinor vanishes on the Killing horizon. In this case, we prove that the spinorial Lie derivative of the Killing spinor with respect to the Killing vector which generates the Killing horizon vanishes, and that all such solutions with more than 13 supersymmetries are pp-waves.}


	\declarations{
		\textbf{Funding.}
		C.S.~has been supported by a Royal Thai Government Doctoral Studentship.\\[5pt]
		\textbf{Conflict of interest.}
		The authors have no relevant financial or non-financial interests to disclose.\\[5pt]
		\textbf{Data statement.}
		No additional research data beyond the data presented and cited in this work are needed to validate the research findings in this work.\\[5pt]
		\textbf{Licence statement.}
		For the purpose of open access, the authors have applied a Creative Commons Attribution (CC-BY) license to any author-accepted manuscript version arising.
	}

	\begin{body}

		\section{Introduction}

		Symmetries of black holes have a particularly important role in our understanding of their properties. Particularly strong classification theorems can be constructed if there are sufficiently many isometries in higher dimensions. For example, in five-dimensional gauged supergravity, it is known that the conditions imposed by supersymmetry are somewhat weaker than in the ungauged theory~\cite{Gauntlett:2002nw,Gauntlett:2003fk}, and it has not yet been possible to construct a general uniqueness theorem for black holes in five-dimensional gauged supergravity, even when supersymmetry is imposed. However, it has been possible to construct supersymmetric black hole uniqueness theorems in this theory if it is assumed that the black holes admit a $T^2$-symmetry~\cite{Lucietti:2023mvj}. Establishing rigidity theorems for black holes is therefore a key starting point towards a systematic classification. Axisymmetry also plays an important role in constructing new solutions, such as the four-dimensional asymptotically de-Sitter black binary solutions in \cite{Dias:2023rde}. The extension of rigidity theorems from their original formulation for asymptotically flat and non-extremal $D=4$ black holes~\cite{Hawking:1971vc,Hawking:1973uf} to higher-dimensional, non-asymptotically flat, and extremal cases is, however, highly non-trivial. For example, the rigidity theorem for extremal higher-dimensional black holes constructed in~\cite{Hollands:2008wn} requires additional `diopantine' conditions in order to hold. A different method for extremal black holes is to establish a rigidity theorem for the near-horizon geometry. Such a construction is done in the near-horizon decoupling limit, and, as such, is insensitive to the asymptotic structure. This has been done in the case of vacuum gravity in any dimension~\cite{Dunajski:2023xrd}, and then extended to include a cosmological constant~\cite{Colling:2024usk}. The most general form of such a horizon rigidity theorem was established in~\cite{Colling:2025dub} for any degenerate horizon satisfying the null energy condition. The challenge for this approach is then to extend the rigidity theorem away from the near-horizon region into the bulk geometry, in a similar way to the analysis of the linear deformation theory of near-horizon extremal Kerr solution in~\cite{Chrusciel:2017vie}.

		In this work, we will utilise supersymmetry to initiate the classification of supersymmetric degenerate Killing horizons in $D=11$ supergravity away from the near-horizon limit. Considerable progress has been made in understanding the properties of near-horizon geometries in this context~\cite{Kunduri:2013gce,Gutowski:2013kma}. In such analysis, there is a natural construction of Gau{\ss}ian null coordinates near the horizon~\cite{Moncrief:1983xua,Kunduri:2013gce}, which we will utilise. For supersymmetric near-horizon geometries, such coordinates also give a particularly useful decomposition of spinors into positive and negative lightcone chirality parts~\cite{Gutowski:2013kma}. This enabled the construction of generalised Lichnerowicz-type theorems for such solutions, which, in turn, implies a doubling of supersymmetry in the near-horizon limit. Such supersymmetry enhancement gives rise to a corresponding symmetry enhancement, and these solutions exhibit (at least) an $\frsl(2,\IR)$-symmetry. Although we do not ordinarily expect such enhancement away from the near-horizon limit, we do expect supersymmetry to impose substantial conditions on supersymmetric Killing horizons \emph{away} from the near-horizon limit, which is the main focus of this work. Moreover, as we will demonstrate, the splitting of the spinors into positive and negative lightcone chirality parts will also be particularly useful away from the near-horizon limit. In addition to establishing general conditions for all supersymmetric Killing horizons away from the near-horizon limit, we also address the issue of highly supersymmetric solutions in the context of Killing horizons.

		The classification of supersymmetric backgrounds in $D=11$ has largely been driven by spinorial geometry~\cite{Gillard:2004xq} and $\sfG$-structure~\cite{Gauntlett:2003wb} methods. For the highly supersymmetric solutions, several prominent results have been established. For example, the maximally supersymmetric backgrounds were fully classified in~\cite{Figueroa-OFarrill:2002ecq}, and it has been shown that any background with $N\geq 30$ supersymmetry is maximally supersymmetric~\cite{Gran:2006cn,Gran:2010tj}, which can be extended to $N>26$ under certain assumptions~\cite{DiBella:2026vyi}. Furthermore, a homogeneity theorem states that any background preserving more than half supersymmetry ($N>16$) is locally homogeneous~\cite{Figueroa-OFarrill:2012kws}.

		Below, we will show that in $D=11$ supergravity every background with $N\geq 14$ supersymmetry that admits a supersymmetric degenerate Killing horizon with a closed spatial cross-section is either a near-horizon geometry or a pp-wave. We begin in \cref{sec:Equations} by introducing the Gau{\ss}ian null coordinates and the basis that we will use throughout the paper. We then lay out the equations for the four-form field strength and the metric in $D=11$ supergravity~\cite{Cremmer:1978km}. In \cref{sec:SloveEquations}, we begin analysing the Killing spinors, deriving results that hold for any number of supersymmetries. We find that, in the case of non-vanishing negative chiral spinors on the horizon, all of the solutions are near-horizon geometries. On the other hand, when the negative chiral spinors vanish on the horizon, we find that every one-form Killing spinor bilinear is proportional to the dual of the Killing vector field which generates the Killing horizon, and the spinorial Lie derivative of the Killing spinor with respect to the Killing vector vanishes. We further analyse the latter case in \cref{sec:N>14Solutions} for $N\geq 14$ supersymmetry. The cases with $N>16$ supersymmetry can be immediately excluded by the homogeneity theorem in~\cite{Figueroa-OFarrill:2012kws}. Moreover, in the case where the Killing vector is timelike, \cref{sec:N>=14Timelike}, we show that there are no solutions with $N \geq 14$ supersymmetry. Furthermore, when the Killing vector is null, \cref{sec:N>=14Null}, we find that every solution with $14 \leq N \leq 16$ supersymmetry is a pp-wave, and it has been shown in~\cite{Hubeny:2002pj,Hubeny:2003ug} that such solutions do not possess event horizons.

		\section{Bosonic sector of \texorpdfstring{$D=11$}{D=11} supergravity}\label{sec:Equations}

		In $D=11$ supergravity, the bosonic fields consist of the metric and the three-form gauge field~\cite{Cremmer:1978km}. In this section, we present the equations governing these fields, then decompose them into lightcone directions and those of the spatial cross-section, where the full expressions are given in the \cref{App:EinsteinEqs,App:GaugeFieldEqs}.

		\subsection{Bases and near-horizon limit}

		Before we present the field equations, we introduce the Gau{\ss}ian-null coordinates and the basis that will be used, along with the near-horizon limit and near-horizon geometry.

		\paragraph{Gau{\ss}ian null coordinates.}
		Since every supersymmetric horizon is a degenerate Killing horizon generated by the dual of the one-form bilinear of the Killing spinor, there exist Gau{\ss}ian null coordinates near the horizon~\cite{Moncrief:1983xua,Kunduri:2013gce}. The metric and four-form field strength in these coordinates $\mu\sim(u,r,y^i)$ with $i,j,\ldots=2,\ldots,10$ take the form
		\begin{subequations}
			\begin{equation}
					g\ =\ \rmd u\odot\big[\rmd r+r\alpha_i(r,y)\rmd y^i-\tfrac12r^2\beta(r,y)\rmd u\big]+\tfrac12\gamma_{ij}(r,y)\rmd y^i\odot\rmd y^j~,
			\end{equation}
			\begin{equation}
				\begin{aligned}
					F\ &=\ \tfrac12F_{urij}(r,y)\rmd u\wedge\rmd r\wedge\rmd y^i\wedge\rmd y^j+\tfrac16F_{rijk}(r,y)\rmd r\wedge\rmd y^i\wedge\rmd y^j\wedge\rmd y^k
					\\
					&\kern1cm+\tfrac16F_{uijk}(r,y)\rmd u\wedge\rmd y^i\wedge\rmd y^j\wedge\rmd y^k+\tfrac{1}{24}F_{ijkl}(r,y)\rmd y^i\wedge\rmd y^j\wedge\rmd y^k\wedge\rmd y^l~,
				\end{aligned}
			\end{equation}
		\end{subequations}
		where $\partial_u$ is the Killing vector field. At fixed $u$ and $r$, which we call the spatial cross-section, $\alpha$, $\beta$, and $\gamma$ are a one-form, a scalar, and a metric, respectively. Throughout this paper, we will assume that the spatial cross-section $S$ is compact without boundary.

		It is well known that $R_{uu}=0$ on a Killing horizon ($r=0$).\footnote{Recall that $R_{uu}=N^{\nu}(\nabla_\mu\nabla_\nu-\nabla_\nu\nabla_\mu)N^\mu$, where $N\coloneqq\partial_u$. Since $N$ is Killing, we have $R_{uu}|_{r=0}=N^{\nu}\nabla_\mu\nabla_\nu N^\mu|_{r=0}=\nabla_\mu N_\nu\nabla^\mu N^\nu|_{r=0}$, where we have used the Killing equation and the fact that $\nabla_NN|_{r=0}=0$. We then use $2\nabla N|_{r=0}=\alpha_i|_{r=0}\rmd r\wedge\rmd y^i$ and conclude that $R_{uu}|_{r=0}=\nabla_\mu N_\nu\nabla^\mu N^\nu|_{r=0}=0$.} Therefore, the $uu$-component of the stress-energy tensor vanishes, which implies that~\cite{Cremmer:1978km,Kunduri:2013gce} (see also~\eqref{eq:CompactEinsteinEq})
		\begin{equation}
			0\ =\ F_{u\mu\nu\rho}F_{u}{}^{\mu\nu\rho}+\caO(r)\ =\ F_{uijk}F_{ulmn}\gamma^{il}\gamma^{jm}\gamma^{kn}+\caO(r)~.
		\end{equation}
		Hence,
		\begin{equation}
			F_{uijk}|_{r=0}\ =\ 0~.
		\end{equation}
		By real analyticity in the $r$-direction, we have
		\begin{subequations}
			\begin{equation}
				F\ \coloneqq\ \rmd u\wedge\rmd r\wedge\Psi+\rmd r\wedge Z+r\rmd u\wedge W+X~,
			\end{equation}
			where 
			\begin{equation}
				\Psi_{ij}\ \coloneqq\ F_{urij}~,
				\quad
				W_{ijk}\ \coloneqq\ \tfrac{1}{r}F_{uijk}~,
				\quad Z_{ijk}\ \coloneqq\ F_{rijk}~,
				\eand
				X_{ijkl}\ \coloneqq\ F_{ijkl}~,
			\end{equation}
		\end{subequations}
		which are all regular at $r=0$.

		\paragraph{Near-horizon limit.}
		The near-horizon limit is the limit $\eps\rightarrow 0$ of a family of diffeomorphisms
		\begin{equation}\label{eq:diffeo}
			(u,r,y^i)\ \rightarrow\ (u/\eps,r\eps,y^i)
			\eforall
			\eps\ >\ 0~.
		\end{equation}
		The metric and the four-form field strength in this limit are
		\begin{subequations}\label{eq:NHGeo}
			\begin{equation}
				\mathring{g}\ \coloneqq\ \rmd u\odot\big[\rmd r+r\mathring{\alpha}_i(y)\rmd y^i-\tfrac12r^2\mathring{\beta}(y)\rmd u\big]+\tfrac12\mathring{\gamma}_{ij}(y)\rmd y^i\odot\rmd y^j
			\end{equation}
			and
			\begin{equation}
				\mathring{F}\ \coloneqq\ \rmd u\wedge\rmd r\wedge\mathring\Psi+\rmd r\wedge\mathring{Z}+r\rmd u\wedge\mathring{W}+\mathring X~,
			\end{equation}
		\end{subequations}
		where $\mathring\alpha\coloneqq\alpha|_{r=0}$, $\mathring\beta\coloneqq\beta|_{r=0}$, $\mathring\gamma\coloneqq\gamma|_{r=0}$, $\mathring\Psi\coloneqq\Psi|_{r=0}$, $\mathring W\coloneqq W|_{r=0}$, $\mathring X\coloneqq X|_{r=0}$, and $\mathring Z\coloneqq Z|_{r=0}$.

		\paragraph{Null orthonormal basis.}
		We will mostly work with a null orthonormal basis, collectively labelled by $\whA\sim(\whpl,\whmi,\wha)$ with $\hat a,\hat b,\ldots=1,\ldots,9$, defined as follows
		\begin{subequations}
			\begin{equation}\label{eq:OrthonormalBasis}
				e^\whpl \ \coloneqq\ \rmd u~,
				\quad
				e^\whmi\ \coloneqq\ -\tfrac12r^2\beta\rmd u+\rmd r +r\alpha_i\rmd y^i~,
				\eand
				e^\wha\ \coloneqq\ e_i{}^\wha\rmd y^i~,
			\end{equation}
			where
			\begin{equation}
				e_i{}^\wha e_j{}^\whb\delta_{\wha\whb}\ =\ \gamma_{ij}~.
			\end{equation}
		\end{subequations}
		In this basis, the metric and the gauge field are given by
		\begin{subequations}
			\begin{equation}\label{eq:MetricOrthonormal}
				g_{\whpl\whpl}\ =\ g_{\whmi\whmi}\ =\ 0~,
				\quad
				g_{\whpl\whmi}\ =\ 1~,
				\eand
				g_{\wha\whb}\ =\ \delta_{\wha\whb}
			\end{equation}
			and
			\begin{equation}\label{eq:GaugeFieldInOrthonormal}
				F\ =\ e^\whpl\wedge e^\whmi\wedge\Psi+r e^\whpl\wedge(W-\alpha\wedge\Psi+\tfrac12r\beta Z)+e^\whmi\wedge Z+X-r\alpha\wedge Z~.
			\end{equation}
		\end{subequations}
		Note that by writing this, we have implicitly defined
		\begin{equation}
			\begin{gathered}
				\alpha_\wha\ \coloneqq\ E_\wha{}^i\alpha_i~,
				\quad
				\Psi_{\wha\whb}\ \coloneqq\ E_\wha{}^iE_\whb{}^j\Psi_{ij}~,
				\quad
				W_{\wha\whb\whc}\ \coloneqq\ E_\wha{}^iE_\whb{}^jE_\whc{}^kW_{ijk}~,
				\\
				X_{\wha\whb\whc\whd}\ \coloneqq\ E_\wha{}^iE_\whb{}^jE_\whc{}^kE_\whd{}^lX_{ijkl}~,
				\eand
				Z_{\wha\whb\whc}\ \coloneqq\ E_\wha{}^iE_\whb{}^jE_\whc{}^kZ_{ijk}~,
			\end{gathered}
		\end{equation}
		where $E_\wha{}^i$ is the inverse of $e_i{}^\wha$.

		\subsection{Bianchi identities and field equations}

		\paragraph{Bianchi Identities.}
		We first consider the Bianchi identities of the gauge field, $\rmd F=0$. The components $urijk$, $uijkl$, $rijkl$, and $ijklm$ of the Bianchi identities give
		\begin{equation}\label{eq:GaugeBianchi}
			\caL_{\partial_r}(rW)\ =\ \tilde\rmd\Psi~,
			\quad
			\tilde\rmd W\ =\ 0~,
			\quad
			\caL_{\partial_r}X\ =\ \tilde\rmd Z~,
			\eand
			\tilde\rmd X\ =\ 0~,
		\end{equation}
		respectively, where `$\tilde\rmd$' is the exterior derivative on the spatial cross-section $S$ and `$\caL_{\partial_r}$' is the Lie derivative with respect to $\partial_r$ which we define to act on any tensor $T$ on the spatial cross-section as
		\begin{equation}
			(\caL_{\partial_r}T)_{i_1\cdots i_n}{}^{j_1\cdots j_m}\ \coloneqq\ \partial_{r}T_{i_1\cdots i_n}{}^{j_1\cdots j_m}
		\end{equation}
		in the Gau{\ss}ian null coordinates. For brevity, we shall denote this Lie derivative also by `$\cdot$' e.g.~$\dot X\coloneqq\caL_{\partial_r}X$ etc.

		\paragraph{Gauge field equation.}
		The four-form field strength satisfies the gauge field equation~\cite{Cremmer:1978km}
		\begin{equation}\label{eq:GaugeEqsShort}
			0\ =\ \nabla^\whA F_{\whA\whB\whC\whD}+\tfrac{1}{2(4!)^2}\varepsilon_{\whB\whC\whD}{}^{\whC_1\cdots\whC_8}F_{\whC_1\cdots\whC_4}F_{\whC_5\cdots\whC_8}~,
		\end{equation}
		where $\varepsilon$ is the volume-form. The expression of the gauge field in components is given in \cref{App:GaugeFieldEqs}.

		\paragraph{Einstein field equation.}
		We will also use the Einstein field equation
		\begin{equation}\label{eq:CompactEinsteinEq}
			R_{\whA\whB}\ =\ \tfrac{1}{12}F_\whA{}^{\whC\whD\whE}F_{\whB\whC\whD\whE}-\tfrac{1}{144}g_{\whA\whB}F_{\whC\whD\whE\whF}F^{\whC\whD\whE\whF}~.
		\end{equation}
		The decomposition of this is given in \cref{App:EinsteinEqs}.

		\section{Killing spinor equation for \texorpdfstring{$D=11$}{D=11} supergravity}\label{sec:SloveEquations}

		Having introduced the bosonic sector, we now consider the fermionic sector and the Killing spinor equation, given by~\cite{Cremmer:1978km}
		\begin{subequations}
			\begin{equation}\label{eq:KSE}
				\nabla_\whA\epsilon+\Big(-\tfrac{1}{288}\Gamma_\whA{}^{\whB\whC\whD\whE}F_{\whB\whC\whD\whE}+\tfrac{1}{36}F_{\whA\whB\whC\whD}\Gamma^{\whB\whC\whD}\Big)\epsilon\ =\ 0~,
			\end{equation}
			where $\epsilon$ is Majorana, and in our notation
			\begin{equation}
				\nabla_\whA\epsilon\ \coloneqq\ E_\whA\epsilon-\tfrac14\omega_{\whA\whB\whC}\Gamma^{\whB\whC}\epsilon~,
			\end{equation}
		\end{subequations}
		where the connection one-form $\omega$ is given in \cref{App:Connection1Form}.

		We define projectors
		\begin{equation}
			\caP_\whpl\ \coloneqq\ \tfrac12\Gamma_\whpl\Gamma_\whmi
			\eand
			\caP_\whmi\ \coloneqq\ \tfrac12\Gamma_\whmi\Gamma_\whpl~.
		\end{equation}
		These decompose the spinor as
		\begin{subequations}
			\begin{equation}
				\epsilon\ =\ \epsilon_\whpl+\epsilon_\whmi~,
			\end{equation}
			where
			\begin{equation}
				\epsilon_{\widehat\pm}\ \coloneqq\ \caP_{\widehat\pm}\epsilon~.
			\end{equation}
		\end{subequations}
		The decomposition of the Killing spinor equation~\eqref{eq:KSE} is given in \cref{App:KillingSpinorEqs}.

		In the remainder of this section, we start solving the Killing spinor equation. We split the calculation into two cases, $\epsilon_\whmi|_{r=0}\neq 0$ and $\epsilon_\whmi|_{r=0}= 0$. We show that if $\epsilon_\whmi|_{r=0}\neq 0$, the solutions are near-horizon solutions~\eqref{eq:NHGeo}. For the case $\epsilon_\whmi|_{r=0}=0$, we show that the one-form Killing spinor bilinears are always proportional to the dual of the Killing vector $\partial_u$, which we analyse further in \cref{sec:N>14Solutions}.

		\subsection{Case \texorpdfstring{$\epsilon_\whmi|_{r=0}\neq 0$}{epsilon-not zero}}

		To proceed in this case, we first consider the Killing spinor equation and the field equations at $r=0$ to derive several identities on the horizon. We then prove by induction that the bosonic fields must take the form of the near-horizon geometry.

		\paragraph{Identities on the horizon.}
		We first observe that by integration parts over the closed spatial cross-section $\mathring S\coloneqq S|_{r=0}$, we have
		\begin{equation}
			\int_{\mathring S}\innerLarge{\Gamma_\whpl\epsilon_\whmi}{\Gamma^\wha\tilde\nabla_\wha\epsilon_\whpl}\Big|_{r=0}\ =\ \int_{\mathring S}\innerLarge{\Gamma_\whpl\Gamma^\wha\tilde\nabla_\wha\epsilon_\whmi}{\epsilon_\whpl}\Big|_{r=0}~,
		\end{equation}
		where $\tilde\nabla$ is the Levi-Civita connection with respect to the metric $\gamma$ on the spatial cross-section $S$ and $\inner{-}{-}$ a Dirac inner product. Using the Killing spinor equation~\eqref{eq:KSEa+} at $r=0$, the left-hand side reads
		\begin{subequations}
			\begin{equation}
				\mathrm{LHS}\ =\ \int_{\mathring S}\innerLarge{\epsilon_\whmi}{\Gamma_\whmi\Big(\tfrac14\alpha_\wha\Gamma^\wha-\tfrac18\Psi_{\wha\whb}\Gamma^{\wha\whb}-\tfrac{1}{96}X_{\wha\whb\whc\whd}\Gamma^{\wha\whb\whc\whd}\Big)\epsilon_\whpl+\tfrac12\dot\gamma_\wha{}^\wha\epsilon_\whmi}\Big|_{r=0}~.
			\end{equation}
			Using the Killing spinor equation~\eqref{eq:KSEa-} at $r=0$, the right-hand side reads
			\begin{equation}
				\mathrm{RHS}\ =\ \int_{\mathring S}\innerLarge{\epsilon_\whmi}{\Gamma_\whmi\Big(\tfrac14\alpha_\wha\Gamma^\wha-\tfrac18\Psi_{\wha\whb}\Gamma^{\wha\whb}-\tfrac{1}{96}X_{\wha\whb\whc\whd}\Gamma^{\wha\whb\whc\whd}\Big)\epsilon_\whpl}\Big|_{r=0}~.
			\end{equation}
		\end{subequations}
		Upon equating these two equations, we have
		\begin{equation}
			\int_{\mathring S}\dot\gamma_\wha{}^\wha\inner{\epsilon_\whmi}{\epsilon_\whmi}\big|_{r=0}\ =\ 0~.
		\end{equation}
		We can use the residual gauge transformation to fix $\dot\gamma_\wha{}^\wha|_{r=0}=\Gamma$ with $\Gamma\in\ker(\mathring{\tilde\nabla}^a\mathring{\tilde\nabla}_a+\mathring\alpha^a\mathring{\tilde\nabla}_a+\mathring{\tilde\nabla}_a\mathring\alpha^a)$~\cite{Gutowski:2025lzi,Fontanella:2016lzo,Dunajski:2023xrd} a constant multiple of a positive definite function~\cite{Dunajski:2023xrd}. This, together with $\epsilon_\whmi\neq0|_{r=0}$, implies that
		\begin{equation}\label{eq:TraceAndEpsilon-id}
			\dot\gamma_\wha{}^\wha\big|_{r=0}\ =\ 0~.
		\end{equation}

		Next, by contracting the Killing spinor equation~\eqref{eq:KSEa+} at $r=0$ with $\Gamma^\wha$ and using~\eqref{eq:TraceAndEpsilon-id}, we get
		\begin{equation}
			\Gamma^\wha\tilde\nabla_\wha\epsilon_\whpl\Big|_{r=0}\ =\ \Big(\tfrac14\alpha_\wha\Gamma^\wha-\tfrac18\Psi_{\wha\whb}\Gamma^{\wha\whb}-\tfrac{1}{96}X_{\wha\whb\whc\whd}\Gamma^{\wha\whb\whc\whd}\Big)\epsilon_\whpl\Big|_{r=0}~,
		\end{equation}
		By the Lichnerowicz-type theorem in~\cite{Gutowski:2013kma}, this equation is equivalent to 
		\begin{equation}
			\tilde\nabla_\wha\epsilon_\whpl\Big|_{r=0}\ =\ \Big(\tfrac14\alpha_\wha-\tfrac{1}{24}\Psi_{\whb\whc}\Gamma_\wha{}^{\whb\whc}+\tfrac16\Psi_{\wha\whb}\Gamma^\whb+\tfrac{1}{288}X_{\whb\whc\whd\whe}\Gamma_\wha{}^{\whb\whc\whd\whe}-\tfrac{1}{36}X_{\wha\whb\whc\whd}\Gamma^{\whb\whc\whd}\Big)\epsilon_\whpl\Big|_{r=0}~.
		\end{equation}
		By substituting this back into the Killing spinor equation~\eqref{eq:KSEa+} at $r=0$, we have
		\begin{equation}\label{eq:KSEa+Atr=0Splitted}
			\Big(-\tfrac14\dot\gamma_{\wha\whb}\Gamma^\whb+\tfrac{1}{72}Z_{\whb\whc\whd}\Gamma_\wha{}^{\whb\whc\whd}-\tfrac{1}{12}Z_{\wha\whb\whc}\Gamma^{\whb\whc}\Big)\epsilon_\whmi\Big|_{r=0}\ =\ 0~.
		\end{equation}

		The next step is to take the inner product of~\eqref{eq:KSEa+Atr=0Splitted} with itself,
		\begin{equation}
			\Big(\tfrac{1}{16}\dot\gamma_{\wha\whb}\dot\gamma^{\wha\whb}+\tfrac{1}{48}Z_{\wha\whb\whc}Z^{\wha\whb\whc}\Big)\inner{\epsilon_\whmi}{\epsilon_\whmi}\Big|_{r=0}\ =\ 0~.
		\end{equation}
		Since $\epsilon_\whmi|_{r=0}\neq 0$, we conclude that
		\begin{equation}
			\dot\gamma|_{r=0}\ =\ 0
			\eand
			Z|_{r=0}\ =\ 0~.
		\end{equation}
		With this, the $\whmi\wha\whb$ components of the gauge field equation~\eqref{eq:GaugeEoM:-b1b2} imply that
		\begin{equation}
			\dot\Psi|_{r=0}\ =\ 0~.
		\end{equation}
		The Bianchi identities~\eqref{eq:GaugeBianchi} also force
		\begin{equation}
			\dot W|_{r=0}\ =\ 0
			\eand
			\dot X|_{r=0}\ =\ 0~.
		\end{equation}

		Finally, we consider the $\whmi\wha$ and $\whpl\whmi$ components of the Einstein field equation~\eqref{eq:EinsteinEq-b} and~\eqref{eq:EinsteinEq+-},
		\begin{subequations}
			\begin{equation}\label{eq:R-aEinstein}
				\begin{aligned}
					0\ &=\ \dot\alpha_\wha-\tfrac12\alpha^\whb\dot\gamma_{\wha\whb}+\tfrac14\alpha_\wha\dot\gamma_\whb{}^\whb-\tfrac12\tilde\nabla_\wha\dot\gamma_\whb{}^\whb+\tfrac12\tilde\nabla^\whb\dot\gamma_{\wha\whb}
					\\
					&\kern1cm+r\Big[\tfrac12\ddot\alpha_\wha+\tfrac12\alpha_\wha\ddot\gamma_\whb{}^\whb-\tfrac12\alpha_\whb\ddot\gamma_\wha{}^\whb-\tfrac12\dot\alpha_\whb\dot\gamma_\wha{}^\whb+\tfrac14\dot\alpha_\wha\dot\gamma_\whb{}^\whb
					\\
					&\kern1cm-\tfrac14\alpha_\wha\dot\gamma^{\whb\whc}\dot\gamma_{\whb\whc}+\tfrac12\alpha_\whb\Big(\dot\gamma^{\whb\whc}\dot\gamma_{\wha\whc}-\tfrac12\dot\gamma_\whc{}^\whc\dot\gamma_\wha{}^\whb\Big)\Big]
					\\
					&\kern1cm-\Big[\tfrac14\Psi_{\whb\whc}Z_\wha{}^{\whb\whc}+\tfrac{1}{12}Z_{\whb\whc\whd}X_\wha{}^{\whb\whc\whd}-\tfrac{r}{12}Z_{\whb\whc\whd}(\alpha\wedge Z)_\wha{}^{\whb\whc\whd}\Big]
				\end{aligned}
			\end{equation}
			and
			\begin{equation}\label{eq:R+-Einstein}
				\begin{aligned}
					0\ &=\ -\beta+\tfrac12\tilde\nabla^\wha\alpha_\wha-\tfrac12\alpha^\wha\alpha_\wha
					\\
					&\kern1cm+r\Big(-2\dot\beta-\tfrac12\beta\dot\gamma_\wha{}^\wha-2\dot\alpha^\wha\alpha_\wha+\tfrac12\alpha_\wha\alpha_\whb\dot\gamma^{\wha\whb}-\tfrac14\alpha_\wha\alpha^\wha\dot\gamma_\whb{}^\whb+\tfrac12\tilde\nabla^\wha\dot\alpha_\wha\Big)
					\\
					&\kern1cm+r^2\Big[-\tfrac12\ddot\beta-\tfrac12\ddot\alpha^\wha\alpha_\wha-\tfrac14\beta \ddot\gamma_\wha{}^\wha-\tfrac14\Big(\dot\beta+\dot\alpha^\whb\alpha_\whb\Big){\dot\gamma}_\wha{}^\wha+\tfrac12\dot\gamma^{\wha\whb}\dot\alpha_\wha\alpha_\whb
					\\
					&\kern1cm-\tfrac12\dot\alpha^\wha\dot\alpha_\wha+\tfrac18\beta\dot\gamma^{\wha\whb}\dot\gamma_{\wha\whb}\Big]
					\\
					&\kern1cm-\Big[-\tfrac{1}{144}X_{\wha\whb\whc\whd}X^{\wha\whb\whc\whd}-\tfrac16\Psi_{\wha\whb}\Psi^{\wha\whb}
					\\
					&\kern1cm+r\Big(\tfrac{1}{18}\alpha_\wha Z_{\whb\whc\whd}X^{\wha\whb\whc\whd}+\tfrac{1}{36}W_{\wha\whb\whc}Z^{\wha\whb\whc}-\tfrac{1}{12}\Psi_{\wha\whb}\alpha_\whc Z^{\wha\whb\whc}\Big)
					\\
					&\kern1cm+r^2\Big(-\tfrac{1}{36}\alpha_\wha\alpha^\wha Z_{\whb\whc\whd}Z^{\whb\whc\whd}+\tfrac{1}{12}\alpha^\wha\alpha_\whd Z_{\wha\whb\whc}Z^{\whb\whc\whd}+\tfrac{1}{72}\beta Z_{\wha\whb\whc}Z^{\wha\whb\whc}\Big)\Big]\,,
				\end{aligned}
			\end{equation}
		\end{subequations}
		where $\tilde\nabla$ is the Levi-Civita connection with respect to the metric $\gamma$ on the spatial cross-section $S$. At $r=0$, these imply that
		\begin{equation}
			\dot\alpha|_{r=0}\ =\ 0
			\eand
			\dot\beta|_{r=0}\ =\ 0~.
		\end{equation}
		
		\paragraph{Inductive proof.}
		Having shown that the $n$-th Lie derivatives $\caL^{(n)}_{\partial_r}\alpha\big|_{r=0}$, $\caL^{(n)}_{\partial_r}\beta\big|_{r=0}$, $\caL^{(n)}_{\partial_r}\gamma\big|_{r=0}$, $\caL^{(n-1)}_{\partial_r}Z\big|_{r=0}$, $\caL^{(n)}_{\partial_r}\Psi\big|_{r=0}$, $\caL^{(n)}_{\partial_r} W\big|_{r=0}$, and $\caL^{(n)}_{\partial_r}X\big|_{r=0}$ vanish for $n=1$, we now prove by induction that they vanish for all $n\geq1$. This result implies that all fields are independent of $r$ in the Gau{\ss}ian null coordinates and every solution is a near-horizon solution.

		To prove this, we assume that $\caL^{(n)}_{\partial_r}\alpha\big|_{r=0}$, $\caL^{(n)}_{\partial_r}\beta\big|_{r=0}$, $\caL^{(n)}_{\partial_r}\gamma\big|_{r=0}$, $\caL^{(n-1)}_{\partial_r}Z\big|_{r=0}$, $\caL^{(n)}_{\partial_r}\Psi\big|_{r=0}$, $\caL^{(n)}_{\partial_r} W\big|_{r=0}$, and $\caL^{(n)}_{\partial_r}X\big|_{r=0}$ vanish for all $1\leq n\leq m$ where $m\in\IN$. We then consider the following algebraic identity
		\begin{subequations}
			\begin{equation}\label{eq:ra+Integrability}
				(\caL_{\partial_r}\tilde\nabla_\wha-\tilde\nabla_\wha\caL_{\partial_r})\epsilon_\whpl\ =\ \tfrac14\tilde\nabla_\whc\dot\gamma_{\wha\whb}\Gamma^{\whb\whc}\epsilon_\whpl~,
			\end{equation}
			where the left-hand side of the above is evaluated using the Killing spinor equation~\eqref{eq:AllKSE} and
			\begin{equation}
				\caL_{\partial_r}\epsilon_{\widehat\pm}\ \coloneqq\ \partial_r\epsilon_{\widehat\pm}-\tfrac14\dot e_\whb{}^\wha\Gamma_\wha{}^\whb\epsilon_{\widehat\pm}~.
			\end{equation}
		\end{subequations}

		We then apply $\caL^{(m-1)}_{\partial_r}$ to~\eqref{eq:ra+Integrability} and evaluate at $r=0$. By the induction hypothesis, the resulting algebraic condition is 
		\begin{equation}
			0\ =\ \Gamma_\whpl\Big[\tfrac14\Big(\caL^{(m+1)}_{\partial_r}\gamma\Big)_{\wha\whb}\Gamma^\whb+\tfrac{1}{12}\Big(\caL^{(m)}_{\partial_r}Z\Big)_{\wha\whb\whc}\Gamma^{\whb\whc}-\tfrac{1}{72}\Big(\caL^{(m)}_{\partial_r}Z\Big)_{\whb\whc\whd}\Gamma_\wha{}^{\whb\whc\whd}\Big]\epsilon_\whmi\Big|_{r=0}~.
		\end{equation}
		This is similar to~\eqref{eq:KSEa+Atr=0Splitted}. Upon taking the inner product with itself and applying the same argument, we obtain
		\begin{equation}\label{eq:recursionGammaAndZResult}
			\caL^{(m+1)}_{\partial_r}\gamma\big|_{r=0}\ =\ 0
			\eand
			\caL^{(m)}_{\partial_r}Z\big|_{r=0}\ =\ 0~.
		\end{equation}

		Next, we apply $\caL^{(m)}_{\partial_r}$ to $\whmi\wha\whb$ components of the gauge field equation~\eqref{eq:GaugeEoM:-b1b2}. After evaluating at $r=0$ and using the induction hypothesis and~\eqref{eq:recursionGammaAndZResult}, we obtain
		\begin{equation}\label{eq:recursionPsiResult}
			\caL^{(m+1)}_{\partial_r}\Psi\big|_{r=0}\ =\ 0~.
		\end{equation}

		Similarly, the Bianchi identities~\eqref{eq:GaugeBianchi} imply
		\begin{equation}\label{eq:recursionWAndXResult}
			\caL^{(m+1)}_{\partial_r}W\big|_{r=0}\ =\ 0
			\eand
			\caL^{(m+1)}_{\partial_r}X\big|_{r=0}\ =\ 0~.
		\end{equation}

		We now apply $\caL^{(m)}_{\partial_r}$ to the $\whmi\wha$ components of the Einstein field equation~\eqref{eq:R-aEinstein} and evaluate at $r=0$. Upon using the induction hypothesis and~\eqref{eq:recursionGammaAndZResult},~\eqref{eq:recursionPsiResult}, and~\eqref{eq:recursionWAndXResult}, we obtain
		\begin{equation}
			0\ =\ \caL^{(m)}_{\partial_r}\dot\alpha\big|_{r=0}+\tfrac12\caL^{(m)}_{\partial_r}(r\ddot\alpha)\big|_{r=0}\ =\ \tfrac{2+m}{2}\caL^{(m+1)}_{\partial_r}\alpha\big|_{r=0}
			\quad\Rightarrow\quad
			\caL^{(m+1)}_{\partial_r}\alpha\big|_{r=0}\ =\ 0~.
		\end{equation}

		Finally, by applying $\caL^{(m+1)}_{\partial_r}$ to the $\whpl\whmi$ component of the Einstein field equation~\eqref{eq:R+-Einstein} and evaluating at $r=0$, we obtain
		\begin{equation}
			0\ =\ -\caL^{(m+1)}_{\partial_r}\beta\big|_{r=0}-2\caL^{(m+1)}_{\partial_r}(r\dot\beta)\big|_{r=0}-\tfrac12\caL^{(m+1)}_{\partial_r}(r^2\ddot\beta)\big|_{r=0}\ =\ \tfrac12(m^2+5m+6)\caL^{(m+1)}_{\partial_r}\beta\big|_{r=0}~.
		\end{equation}

		Altogether, by induction, we have proved that $\caL_{\partial_r}^{(n)}$ acting on $\alpha$, $\beta$, $\gamma$, $\Psi$, $W$, $X$, and $Z$ vanishes at $r=0$ for all $n\geq 1$ and $Z|_{r=0}=0$. Hence, every solution is a near-horizon solution.

		\subsection{Case \texorpdfstring{$\epsilon_\whmi|_{r=0}=0$}{epsilon-zero}}

		For this case, we will show that the spinors are independent of $u$ and that every one-form spinor bilinear is proportional to the dual of the Killing vector field $\partial_u$. We leave the full analysis of the Killing spinor and the field equations to the next section.

		\paragraph{$u$-independence.}
		Instead of showing $\partial_u\epsilon_{\widehat\pm}=0$ directly, we show by induction that, equivalently, $\partial_u\caL_{\partial_r}^{(n)}\epsilon_{\widehat\pm}\big|_{r=0}=0$ for all integers $n\geq0$, which implies that $\partial_u\epsilon_{\widehat\pm}=0$.\footnote{Note that $\partial_u$ commutes with $\caL_{\partial_r}$ since the bosonic fields are independent of $u$} This is true for $n=0$ due to the Killing spinor equation~\eqref{eq:KSEu+} at $r=0$ and $\epsilon_\whmi\big|_{r=0}=0$. We then assume that $\partial_u\caL_{\partial_r}^{(n)}\epsilon_{\widehat\pm}\big|_{r=0}=0$ for all $0\leq n \leq m $ where $m\in\IN$. By applying $\partial_u\caL_{\partial_r}^{(m)}$ to the Killing spinor equations~\eqref{eq:KSEr+} and~\eqref{eq:KSEr-}, we obtain
		 \begin{subequations}
			\begin{equation}
				\partial_u\caL_{\partial_r}^{(m+1)}\epsilon_\whpl\ =\ -\tfrac{1}{24}\caL_{\partial_r}^{(m)}\Big(Z_{\wha\whb\whc}\Gamma^{\wha\whb\whc}\partial_u\epsilon_\whpl\Big)
			\end{equation}
			and
			\begin{equation}
				\partial_u\caL_{\partial_r}^{(m+1)}\epsilon_\whmi\ =\ \caL_{\partial_r}^{(m)}\Big(\Gamma_\whmi\Theta_\whpl\partial_u\epsilon_\whpl-\tfrac{1}{72}Z_{\wha\whb\whc}\Gamma^{\wha\whb\whc}\partial_u\epsilon_\whmi\Big)\,.
			\end{equation}
		\end{subequations}
		The right-hand sides of both equations vanish at $r=0$ by the induction hypothesis. Hence, $\partial_u\caL_{\partial_r}^{(m+1)}\epsilon_{\widehat\pm}\big|_{r=0}=0$ and so, we may conclude that $\partial_u\caL_{\partial_r}^{(n)}\epsilon_{\widehat\pm}\big|_{r=0}=0$ for all $n\geq 0$. Therefore,
		\begin{equation}\label{eq:uIndependenceOfSpinor}
			\partial_u\epsilon_{\widehat\pm}\ =\ 0~.
		\end{equation}

		\paragraph{Killing spinor bilinears.}
		We will now consider spinor bilinears and show that every one-form Killing spinor bilinear is proportional to the dual of the vector field $\partial_u$. We first define the space of Killing spinors as $\spn_\IR\{\epsilon^{(i)}\,|\,i=1,\ldots,N\}$, where $N$ is the number of supersymmetries.

		Next, we introduce one-form bilinears as~\cite{Gran:2005wu} (see also \cref{App:SpinorForm})
		\begin{equation}
			K_\whA^{(ij)}\ \coloneqq\ \tfrac{1}{\sqrt{2}}\innerLarge{(\Gamma_\whpl-\Gamma_\whmi)\epsilon^{(i)}}{\Gamma_\whA\epsilon^{(j)}}.
		\end{equation}
		Explicitly,
		\begin{equation}
			\begin{aligned}
				K_\whpl^{(ij)}\ &=\ \sqrt{2}\innerLarge{\epsilon_\whmi^{(i)}}{\epsilon_\whmi^{(j)}},
				\\
				K_\whmi^{(ij)}\ &=\ -\sqrt{2}\innerLarge{\epsilon_\whpl^{(i)}}{\epsilon_\whpl^{(j)}},
				\\
				K_\wha^{(ij)}\ &=\ \tfrac{1}{\sqrt{2}}\Big(\innerLarge{\epsilon_\whmi^{(i)}}{\Gamma_\whmi\Gamma_\wha\epsilon_\whpl^{(j)}}+\innerLarge{\epsilon_\whmi^{(j)}}{\Gamma_\whmi\Gamma_\wha\epsilon_\whpl^{(i)}}\Big)\,.
			\end{aligned}
		\end{equation}

		We first note that the Killing spinor equation~\eqref{eq:KSEr+} implies that
		\begin{equation}\label{eq:EpsilinPlusijrIndependence}
			\begin{aligned}
				\partial_r\innerLarge{\epsilon^{(i)}_\whpl}{\epsilon^{(j)}_\whpl}\ &=\ \innerLarge{\Big(\tfrac14\dot e_\whb{}^\wha\Gamma_\wha{}^\whb-\tfrac{1}{24}Z_{\wha\whb\whc}\Gamma^{\wha\whb\whc}\Big)\epsilon_\whpl^{(i)}}{\epsilon^{(j)}_\whpl}
				\\
				&\kern1cm+\innerLarge{\epsilon^{(i)}_\whpl}{\Big(\tfrac14\dot e_\whb{}^\wha\Gamma_\wha{}^\whb-\tfrac{1}{24}Z_{\wha\whb\whc}\Gamma^{\wha\whb\whc}\Big)\epsilon_\whpl^{(j)}}\ =\ 0~.
			\end{aligned}
		\end{equation}

		Using the Killing spinor equations~\eqref{eq:KSEr-},~\eqref{eq:KSEu+},~\eqref{eq:KSEa+}, and~\eqref{eq:uIndependenceOfSpinor} and the field equations in \cref{App:GaugeFieldEqs,App:EinsteinEqs}, we then compute
		\begin{equation}
			\begin{aligned}
				&\mathring{\tilde\nabla}^\wha\mathring{\tilde\nabla}_\wha\innerLarge{\epsilon_\whpl^{(i)}}{\epsilon_\whpl^{(j)}}\Big|_{r=0}-\mathring\alpha^\wha\mathring{\tilde\nabla}_\wha\innerLarge{\epsilon_\whpl^{(i)}}{\epsilon_\whpl^{(j)}}\Big|_{r=0}
				\\
				&\kern1cm=\ \innerLarge{\epsilon_\whpl^{(i)}}{\Big[\mathring\beta-2\big(\mathring{\bar\Theta}_\whmi\mathring\Theta_\whpl+\mathring\Theta_\whmi\mathring{\bar\Theta}_\whpl\Big)\Big]\epsilon_\whpl^{(j)}}\Big|_{r=0}
				\\
				&\kern1cm=\ -\innerLarge{\epsilon_\whpl^{(i)}}{\partial_r\partial_u\epsilon_\whpl^{(j)}}\Big|_{r=0}-\innerLarge{\partial_r\partial_u\epsilon_\whpl^{(i)}}{\epsilon_\whpl^{(j)}}\Big|_{r=0}\ =\ 0~.
			\end{aligned}
		\end{equation}
		Note that $\mathring{\bar\Theta}_{\widehat\pm}$ is the near-horizon limit of $\bar\Theta_{\widehat\pm}$ defined in~\eqref{eq:DefTheta}. Since the spatial cross-section is compact without boundary, $\inner{\epsilon_\whpl^{(i)}}{\epsilon_\whpl^{(j)}}\big|_{r=0}$ is constant by the maximum principle~\cite{PUCCI20041}. Using this result,~\eqref{eq:EpsilinPlusijrIndependence} then implies that $\inner{\epsilon_\whpl^{(i)}}{\epsilon_\whpl^{(j)}}$ must be constant everywhere.

		Next, we compute
		\begin{equation}
			\begin{aligned}
				&\caL_{\partial_r}\Big(\innerLarge{\epsilon_\whmi^{(i)}}{\Gamma_\whmi\Gamma^\wha\epsilon_\whpl^{(j)}}+\innerLarge{\epsilon_\whmi^{(j)}}{\Gamma_\whmi\Gamma^\wha\epsilon_\whpl^{(i)}}\Big)
				\\
				&\kern1cm=\ \partial_r\innerLarge{\epsilon_\whmi^{(i)}}{\Gamma_\whmi\Gamma^\wha\epsilon_\whpl^{(j)}}+\partial_r\innerLarge{\epsilon_\whmi^{(j)}}{\Gamma_\whmi\Gamma^\wha\epsilon_\whpl^{(i)}}
				\\
				&\kern2cm-\dot e_\whb{}^\wha\innerLarge{\epsilon_\whmi^{(i)}}{\Gamma_\whmi\Gamma^\whb\epsilon_\whpl^{(j)}}-\dot e_\whb{}^\wha\innerLarge{\epsilon_\whmi^{(j)}}{\Gamma_\whmi\Gamma^\whb\epsilon_\whpl^{(i)}}
				\\
				&\kern1cm=\ \innerLarge{\epsilon_\whpl^{(i)}}{2(\Theta_-\Gamma^\wha+\Gamma^\wha\Theta_+)\epsilon_\whpl^{(j)}}
				\\
				&\kern2cm+\innerLarge{\epsilon_\whmi^{(i)}}{\Gamma_\whmi\Big(-\tfrac12\dot\gamma^{\wha\whb}\Gamma_\whb-\tfrac{1}{36}Z_{\whb\whc\whd}\Gamma^{\wha\whb\whc\whd}-\tfrac16Z^{\wha\whb\whc}\Gamma_{\whb\whc}\Big)\epsilon_\whpl^{(j)}}
				\\
				&\kern2cm+\innerLarge{\epsilon_\whmi^{(j)}}{\Gamma_\whmi\Big(-\tfrac12\dot\gamma^{\wha\whb}\Gamma_\whb-\tfrac{1}{36}Z_{\whb\whc\whd}\Gamma^{\wha\whb\whc\whd}-\tfrac16Z^{\wha\whb\whc}\Gamma_{\whb\whc}\Big)\epsilon_\whpl^{(i)}}
				\\
				&\kern1cm=\ 2\tilde\nabla^\wha\innerLarge{\epsilon_\whpl^{(i)}}{\epsilon_\whpl^{(j)}}\ =\ 0~,
			\end{aligned}
		\end{equation}
		where the last equality follows from the constancy of $\inner{\epsilon_\whpl^{(i)}}{\epsilon_\whpl^{(j)}}$. Therefore, by the initial condition $\epsilon_\whmi|_{r=0}=0$, we arrive at
		\begin{equation}\label{eq:(mi,miapl)=0N>1}
			\sqrt{2}K_\wha^{(ij)}\ =\ \innerLarge{\epsilon_\whmi^{(i)}}{\Gamma_\whmi\Gamma_\wha\epsilon_\whpl^{(j)}}+\innerLarge{\epsilon_\whmi^{(j)}}{\Gamma_\whmi\Gamma_\wha\epsilon_\whpl^{(i)}}\ =\ 0~.
		\end{equation}

		Finally, using~\eqref{eq:(mi,miapl)=0N>1}, we compute
		\begin{equation}
			\begin{aligned}
				\caL_{\partial_r}\innerLarge{\epsilon_\whmi^{(i)}}{\epsilon_\whmi^{(j)}}\ &=\ \innerLarge{\epsilon_\whmi^{(i)}}{\Gamma_\whmi\Theta_\whpl\epsilon_\whpl^{(j)}}+\innerLarge{\epsilon_\whpl^{(i)}}{\Gamma_\whpl\bar\Theta_\whmi\epsilon_\whmi^{(j)}}
				\\
				&=\ \partial_u\innerLarge{\epsilon_\whpl^{(i)}}{\epsilon_\whpl^{(j)}}+\big(r\beta+\tfrac12 r^2\dot\beta\big)\innerLarge{\epsilon_\whpl^{(i)}}{\epsilon_\whpl^{(j)}}
				\\
				&\kern1cm-\tfrac12\big(\alpha^\wha+r\dot\alpha^\wha\big)\Big(\innerLarge{\epsilon_\whpl^{(i)}}{\Gamma_\whpl\Gamma_\wha\epsilon_\whmi^{(j)}}+\innerLarge{\epsilon_\whpl^{(j)}}{\Gamma_\whpl\Gamma_\wha\epsilon_\whmi^{(i)}}\Big)
				\\
				&=\ \caL_{\partial_r}\Big(\tfrac12r^2\beta\innerLarge{\epsilon_\whpl^{(i)}}{\epsilon_\whpl^{(j)}}\Big)\,.
			\end{aligned}
		\end{equation}
		Therefore, by the initial condition $\epsilon_\whmi|_{r=0}=0$,
		\begin{equation}\label{eq:Relation(Pl,Pl)(Mi,Mi)N>1}
			\innerLarge{\epsilon_\whmi^{(i)}}{\epsilon_\whmi^{(j)}}\ =\ \tfrac12r^2\beta\innerLarge{\epsilon_\whpl^{(i)}}{\epsilon_\whpl^{(j)}}\,.
		\end{equation}
		
		Collecting all of the results, we may conclude that
		\begin{equation}\label{eq:BilinearMatching}
			K^{(ij)}\ =\ C^{(ij)}\big(-\tfrac12r^2\beta e^\whpl+e^\whmi\big)
			\ewith
			C^{(ij)}\ \coloneqq\ -\sqrt{2}\innerLarge{\epsilon^{(i)}_\whpl}{\epsilon_\whpl^{(j)}}
		\end{equation}
		and $C^{(ij)}$ is a constant for each $i$ and $j$. Note that the one-form in the bracket is simply the dual of the vector field $\partial_u$.

		\section{Solutions with \texorpdfstring{$N\geq14$}{N>=14} bulk supersymmetry and \texorpdfstring{$\epsilon_\whmi|_{r=0}=0$}{epsilon-zero}}\label{sec:N>14Solutions}

		Having solved the case $\epsilon_\whmi|_{r=0}\neq0$, we now consider the case when $\epsilon_\whmi|_{r=0}=0$. We first remark that solutions with $N>16$ bulk supersymmetry can immediately be excluded. This is because all one-form spinor bilinears $K$ are proportional to the same single one-form~\eqref{eq:BilinearMatching} in this case, and it is known from~\cite{Figueroa-OFarrill:2012kws} that the set of all such bilinears (for $N>16$) pointwise spans the space of all one-forms, which is clearly not the case here. Hence, $N>16$ backgrounds are excluded in this case. 

		In the remainder of this section, we show that solutions with $N\geq14$ supersymmetry and $\epsilon_\whmi|_{r=0}=0$ are pp-waves. Explicitly, we prove that there are no solutions with $\beta\neq0$, while for the case $\beta=0$, the solutions are pp-waves, which do not possess an event horizon~\cite{Hubeny:2002pj,Hubeny:2003ug}. In these calculations, we rely heavily on the spinorial geometry conventions set out in~\cite{Gran:2005wu}.

		\subsection{Non-existence of \texorpdfstring{$N\geq14$}{N>=14} solutions with \texorpdfstring{$\beta\neq0$}{beta!=0}}\label{sec:N>=14Timelike}

		We begin by considering the case in which $\partial_u$ is timelike in some open set $U$ (i.e.~$\beta>0$ on $U$)\footnote{Note that~\eqref{eq:Relation(Pl,Pl)(Mi,Mi)N>1} implies that $\beta\geq0$. Therefore, if $\beta\neq0$, such an open set $U$ exists.}, and, furthermore, consider an open set $W\subseteq U$ in which the solution preserves $N\geq14$ supersymmetry. We let $p\in W$ be a fixed but arbitrary point, and from now on, in this subsection, all identities are evaluated pointwise at $p$.

		\paragraph{One-form bilinears.}
		Instead of the null orthonormal basis~\eqref{eq:OrthonormalBasis}, for convenience, we shall now work in an orthonormal basis labelled by indices $\ulA\sim(\underline{0},\ula)$ with $\ula,\ulb,\ldots=\underline{1},\ldots,\underline{10}$ in which the components of the metric are $g_{\ulA\ulB}=\eta_{\ulA\ulB}$ with $(\eta_{\ulA\ulB})=\diag(-1,1,\ldots,1)$. In addition, we shall represent spinors as differential forms on complexified $\IR^5$ and work with the timelike representation of the gamma matrices as summarised in \cref{App:SpinorForm} (see also~\cite[Appendix A.1]{Gran:2005wu}).

		In particular, an arbitrary Majorana spinor takes the form
		\begin{equation}
			\epsilon\ =\ \rho\cdot1+\bar\rho\bfe_{12345}+\chi^m\bfe_m +\tfrac{1}{4!}({\star\bar\chi})^{mnop}\bfe_{mnop}+\tfrac12\sigma^{mn}\bfe_{mn}-\tfrac{1}{3!}({\star\bar\sigma})^{mno}\bfe_{mno}~,
		\end{equation}
		where $1$, $\bfe_m$, $\bfe_{mn}$, etc.~generate $\Lambda^\bullet\IR^5$ with $m,n,\ldots=1,\ldots,5$, `$\star$' denotes the Hodge dual with respect to the standard Euclidean metric on $\IR^5$, and $\rho$, $\chi^m$, and $\sigma^{mn}$ are in general complex-valued. We remark that if $\epsilon$ and $\eta$ are Majorana spinors then the one-form bilinear defined in~\eqref{eq:timelikeOneFormBilinear} satisfies
		\begin{equation}
			B(\epsilon,\Gamma_\ulA\eta)\ =\ B(\eta,\Gamma_\ulA\epsilon)
		\end{equation}
		and so, if the space of Killing spinors is $\{\epsilon^{(i)}\,|\,i=1,\ldots,N\}$, where $N$ is the number of supersymmetries, then~\eqref{eq:BilinearMatching} implies
		\begin{equation}
			B(\epsilon^{(i)},\Gamma_\ulA\epsilon^{(j)})\ =\ C^{(ij)}V_\ulA~,
		\end{equation}
		where $C^{(ij)}$ is real symmetric and $V_\ulA$ is a real one-form. Without loss of generality, we apply an $\sfSO(N)$-rotation to the space of spinors such that $C^{(ij)}$ is diagonalised. In such a basis, we have
		\begin{equation}\label{eq:inprod1}
			B(\epsilon^{(i)},\Gamma_\ulA\epsilon^{(j)})\ =\ \kappa^{(i)}\delta^{(ij)}V_\ulA
		\end{equation}
		with $\kappa^{(i)}$ the eigenvalues. On evaluating this identity when $\ulA=\underline{0}$ and $i=j$, we find
		\begin{equation}
			-\innerLarge{\Gamma_{\underline{0}}\epsilon^{(i)}}{\Gamma_{\underline{0}}\epsilon^{(i)}}\ =\ \kappa^{(i)}V_{\underline{0}}~,
		\end{equation}
		where we used the definition of $B$ and the Majorana condition in \cref{App:SpinorForm}. This implies that $\kappa^{(i)}\neq0$ and $V_{\underline{0}}\neq0$ (and so, without loss of generality, we can normalise the spinors such that $\kappa^{(i)}=\pm 1$). Furthermore, on considering the one-form identity~\eqref{eq:inprod1} for the case $\ulA=\underline{0}$ and $i\neq j$, we find
		\begin{equation}
			\innerLarge{\epsilon^{(i)}}{\epsilon^{(j)}}\ =\ 0~.
		\end{equation}

		\paragraph{First spinor in canonical form.}
		To proceed, we note that a $\sfSpin(1,10)$-transformation can be used to set
		\begin{equation}\label{eq:firstSpinor}
			\epsilon^{(1)}\ =\ f(1+\bfe_{12345})
		\end{equation}
		for $f\in\IR\setminus\{0\}$~\cite{Gillard:2004xq}. Upon substituting this $\epsilon^{(1)}$ into~\eqref{eq:inprod1}, we find that $V_\ula=0$ and $V_{\underline{0}}\neq0$. 

		\paragraph{Second spinor in canonical form.}
		To construct $\epsilon^{(2)}$, we set
		\begin{equation}
			\epsilon^{(2)}\ =\ \rho\cdot 1+\bar\rho\bfe_{12345}+\chi^m\bfe_m+\tfrac{1}{4!}({\star\bar\chi})^{mnop}\bfe_{mnop}+\tfrac12\sigma^{mn}\bfe_{mn}-\tfrac{1}{3!}({\star\bar\sigma})^{mno}\bfe_{mno}~.
		\end{equation}
		Upon imposing the condition~\eqref{eq:inprod1}, we have
		\begin{equation}
			B(1+\bfe_{12345},\Gamma_\ulA\epsilon^{(2)})\ =\ 0
		\end{equation}
		and so,
		\begin{equation}\label{eq:cond1}
			\rho+\bar\rho\ =\ 0
			\eand
			\chi^m\ =\ 0
			\eforall
			m\ =\ 1,\ldots,5~.
		\end{equation}
		In particular, since all the $\chi^m$ are forced to be zero, it follows from the reasoning given in~\cite[Appendix A]{Gillard:2004xq} that an $\sfSU(5)$-transformation can be used, which leaves $\epsilon^{(1)}$ invariant, to write $\epsilon^{(2)}$ as
		\begin{equation}
			\epsilon^{(2)}\ =\ \rmi x(1-\bfe_{12345})+\sigma^{12}\big(\bfe_{12}-\bfe_{345})+\sigma^{34}(\bfe_{34}-\bfe_{125})
		\end{equation}
		for $x,\sigma^{12},\sigma^{34}\in\IR$ and `i' the imaginary unit. Again, the condition~\eqref{eq:inprod1} implies that
		\begin{equation}
			B(\epsilon^{(2)},\Gamma_\ula\epsilon^{(2)})\ =\ 0
		\end{equation}
		and so,
		\begin{equation}
			\sigma^{12}\sigma^{34}\ =\ 0~.
		\end{equation}
		Now, as $\rmi x(1-\bfe_{12345})+k_1(\bfe_{12}-\bfe_{345})$ and $\rmi x(1-\bfe_{12345})+k_2(\bfe_{34}-\bfe_{125})$ are $\sfSU(5)$ gauge-equivalent (for $x,k_1,k_2\in\IR$), it follows that without loss of generality we may take
		\begin{equation}\label{eq:simp2form}
			\epsilon^{(2)}\ =\ \rmi x(1-\bfe_{12345})+\sigma^{12}\big(\bfe_{12}-\bfe_{345})
		\end{equation}
		for $x,\sigma^{12}\in\IR$.
		
		Suppose first that $\sigma^{12}=0$ (and so, $x\neq 0$). Then, consider $\tilde\epsilon^{(3)}$. As both $\epsilon^{(1)}$ and $\epsilon^{(2)}$ are invariant under $\sfSU(5)$, it follows that, using the same reasoning as given above, one can write
		\begin{equation}
			\tilde\epsilon^{(3)}\ =\ k_3(\bfe_{12}-\bfe_{345})~,
		\end{equation}
		where we note that the orthogonality condition with $\epsilon^{(1)}$ and $\epsilon^{(2)}$ sets the components of $1$ and $\bfe_{12345}$ in $\tilde\epsilon^{(3)}$ to be zero. This expression for $\tilde\epsilon^{(3)}$ is identical to that given in~\eqref{eq:simp2form} but with $x=0$.

		Consequently, without loss of generality, we can take $\epsilon^{(2)}$ to be given by
		\begin{equation}\label{eq:secondSpinor}
			\epsilon^{(2)}\ =\ \rmi x(1-\bfe_{12345})+k(\bfe_{12}-\bfe_{345})
		\end{equation}
		with $x,k\in\IR$ and $k\neq0$.

		\paragraph{Remaining spinors.}
		Now consider the remaining $N-2$ Killing spinors $\epsilon^{(j)}$ for $j=3,\ldots,N$. If we take an arbitrary spinor in this set, $\epsilon^{(j)}$, and write again
		\begin{equation}
			\epsilon^{(j)}\ =\ \rho\cdot 1+\bar\rho \bfe_{12345}+\chi^m\bfe_m+\tfrac{1}{4!}({\star\bar\chi})^{mnop}\bfe_{mnop}+\tfrac12\sigma^{mn}\bfe_{mn}-\tfrac{1}{3!}({\star\bar\sigma})^{mno}\bfe_{mno}~,
		\end{equation}
		then again the condition~\eqref{eq:inprod1} yields
		\begin{equation}
			B(1+\bfe_{12345},\Gamma_\ulA\epsilon^{(j)})\ =\ 0
		\end{equation}
		and so,
		\begin{equation}\label{eq:ccondsp1}
			\rho\ =\ \rmi y
			\eand
			\chi^m\ =\ 0
			\eforall
			m\ =\ 1,\ldots,5
		\end{equation}
		for $y\in\IR$ as for $\epsilon^{(2)}$ in our discussion above. Furthermore, the condition~\eqref{eq:inprod1} yields
		\begin{equation}
			B(\rmi x(1-\bfe_{12345})+k(\bfe_{12}-\bfe_{345}),\Gamma_\ulA\epsilon^{(j)})\ =\ 0
		\end{equation}
		and so,
		\begin{subequations}
			\begin{equation}\label{eq:ccondsp2}
				\sigma^{34}\ =\ \sigma^{35}\ =\ \sigma^{45}\ =\ 0~,
			\end{equation}
			as well as
			\begin{equation}\label{eq:ccondsp3}
				-k(\sigma^{12}+{\bar{\sigma}^{12}})+2xy\ =\ 0~.
			\end{equation}
		\end{subequations}
		If we set
		\begin{equation}
			\begin{aligned}
				\caS_{14}\ &\coloneqq\ \spn_\IR\big\{\bfe_{13}+\bfe_{245},\rmi(\bfe_{13}-\bfe_{245}),\bfe_{14}-\bfe_{235},\rmi(\bfe_{14}+\bfe_{235}),\bfe_{15}+\bfe_{234},\rmi(\bfe_{15}-\bfe_{234}),
				\\
				&\kern1cm\bfe_{23}-\bfe_{145},\rmi(\bfe_{23}+\bfe_{145}),\bfe_{24}+\bfe_{135},\rmi(\bfe_{24}-\bfe_{135}),\bfe_{25}-\bfe_{134},\rmi(\bfe_{25}+\bfe_{134})
				\\
				&\kern1cm\rmi(\bfe_{12}+\bfe_{345}),\tfrac{x}{k}(\bfe_{12}-\bfe_{345})+\rmi(1-\bfe_{12345})\big\}\,,
			\end{aligned}
		\end{equation}
		it follows that the conditions~\eqref{eq:ccondsp1},~\eqref{eq:ccondsp2}, and \eqref{eq:ccondsp3} imply that $\epsilon^{(j)}\in\caS_{14}$ for all $j=3,\ldots,N$.

		In particular, for $N\geq 14$, that is, $N-2\geq12$ and, consequently, there must exist a Killing spinor $\epsilon^{(3)}$ of the form
		\begin{equation}
			\begin{aligned}
				\epsilon^{(3)}\ &=\ \sigma^{13}\bfe_{13}+\bar\sigma^{13}\bfe_{245}+\sigma^{14}\bfe_{14}-\bar\sigma^{14}\bfe_{235}+\sigma^{15}\bfe_{15}+\bar\sigma^{15}\bfe_{234}
				\\
				&\kern1cm+\sigma^{23}\bfe_{23}-\bar\sigma^{23}\bfe_{145}+\sigma^{24}\bfe_{24}+\bar\sigma^{24}\bfe_{135}+\sigma^{25}\bfe_{25}-\bar\sigma^{25}\bfe_{134}
				\\
				&\kern1cm+\rmi q_1(\bfe_{12}+\bfe_{345})+q_2\big[\tfrac{x}{k}(\bfe_{12}-\bfe_{345})+\rmi(1-\bfe_{12345})\big]
			\end{aligned}
		\end{equation}
		for $\sigma^{13}$, $\sigma^{14}$, $\sigma^{15}$, $\sigma^{23}$, $\sigma^{24}$, $\sigma^{25}\in\IC$, and $q_1,q_2\in\IR$, and for which not all of $\{\sigma^{13},\sigma^{14},\sigma^{15}\}$ are zero. We can then simplify $\epsilon^{(3)}$ considerably by first applying an $\sfSU(3)$-transformation which leaves invariant $\epsilon^{(1)}$ (see~\eqref{eq:firstSpinor}) and $\epsilon^{(2)}$ (see~\eqref{eq:secondSpinor}) to set without loss of generality $\sigma^{14}=0$ and $\sigma^{15}=0$. Then, the condition~\eqref{eq:inprod1} yields
		\begin{equation}\label{eq:ConstaintsOncaV}
			B(\epsilon^{(3)},\Gamma_\ula\epsilon^{(3)})\ =\ 0
		\end{equation}
		which, in turn, implies that $\sigma^{13}\sigma^{25}=\sigma^{13}\sigma^{24}=0$. Hence, $\sigma^{25}=0$ and $\sigma^{24}=0$. One can then apply an $\sfSU(2)$-transformation which leaves $\epsilon^{(1)}$ and $\epsilon^{(2)}$ invariant to then further set, without loss of generality $\sigma^{23}=0$ and $\sigma^{13}\in\IR\setminus\{0\}$ giving
		\begin{equation}
			\epsilon^{(3)}\ =\ \sigma^{13}(\bfe_{13}+\bfe_{245})+\rmi q_1(\bfe_{12}+\bfe_{345})+q_2\big[\tfrac{x}{k}(\bfe_{12}-\bfe_{345})+\rmi(1-\bfe_{12345})\big] \ .
		\end{equation}

		To proceed further, consider the set of 23 spinors given by
		\begin{equation}
			\begin{aligned}
				\caT\ &\coloneqq\ \big\{\epsilon^{(1)},\epsilon^{(2)},\epsilon^{(3)}\big\}\ \cup\ \big\{B\Gamma_\ula\epsilon^{(1)}\,\big|\,1\leq\ula\leq10,~\ula\notin\{1,6\}\big\}
				\\
				&\kern1cm\cup\ \big\{B\Gamma_\ula\epsilon^{(2)}\,\big|\,1\leq\ula\leq10,~\ula\notin\{1,6\}\big\}\ \cup\ \big\{B\Gamma_\ula\epsilon^{(3)}\,\big|\,\ula \in\{4,5,9,10\}\big\}\,.
			\end{aligned}
		\end{equation}
		Note that~\eqref{eq:inprod1} implies that the elements $\{\epsilon^{(1)},\epsilon^{(2)},\epsilon^{(3)}\}$ are orthogonal with respect to the Dirac inner product to the remaining 20 elements of $\caT$, and, moreover, the set of 20 spinors
		\begin{equation}
			\big\{\Gamma_\ula\epsilon^{(1)}\,\big|\,1\leq\ula\leq10,\ula\notin\{1,6\}\big\}\ \cup\ \big\{\Gamma_\ula\epsilon^{(2)}\,\big|\,1\leq\ula\leq10,\ula\notin\{1,6\}\big\}\ \cup\ \big\{\Gamma_\ula\epsilon^{(3)}\,\big|\,\ula\in\{4,5,9,10\}\big\}
		\end{equation}
		is itself linearly independent. This can be shown by constructing the $20\times32$ matrix whose components correspond to the above components of the above 20 spinors in the standard basis. 12 of the rows of this matrix are identically zero, and on eliminating these, the determinant of the resulting $20\times20$ matrix is equal to $1024f^8k^8(\sigma^{13})^4\neq0$. Consequently, $\caT$ contains 23 linearly independent spinors, and all of these spinors are orthogonal with respect to the Dirac inner product to the set $\{\epsilon^{(j)}\,|\,4\leq j\leq N\}$ of $N-3$ Killing spinors as a consequence of~\eqref{eq:inprod1}. It therefore follows that $N-3\leq9$ and, consequently, $N\leq12$. In conclusion, this implies that $N\geq14$ supersymmetry is not possible if $\beta\neq0$.

		\subsection{\texorpdfstring{$N\geq14$}{N>=14} solutions with \texorpdfstring{$\beta=0$}{beta=0}}\label{sec:N>=14Null}

		Having excluded $N\geq14$ solutions with $\beta\neq0$, we now consider the special case of $\beta=0$. To this end, we shall work again in the null orthonormal basis~\eqref{eq:OrthonormalBasis}. We first note that since $\beta=0$,~\eqref{eq:Relation(Pl,Pl)(Mi,Mi)N>1} implies that $\epsilon_\whmi=0$, and, consequently, for $N\geq14$ spinors, the space of $\epsilon_\whpl$ spinors is an $(N\geq14)$-dimensional space of positive chirality spinors.

		\paragraph{Identities from homogeneity.}
		To proceed, we first derive some identities using a similar argument as the homogeneity theorem in~\cite{Figueroa-OFarrill:2012kws}. Since $\epsilon_\whmi=0$, the Killing spinor equations~\eqref{eq:KSEr-},~\eqref{eq:KSEu+}, and~\eqref{eq:KSEa-} simplify to
		\begin{subequations}
			\begin{equation}
				0\ =\ \Theta_\whpl\epsilon_\whpl
			\end{equation}
			and
			\begin{equation}\label{eq:KSEu+N>8Beta=0}
				0\ =\ \Big[\tfrac18\Big(r(\tilde\rmd\alpha)_{\wha\whb}+2r^2\dot\alpha_\wha\alpha_\whb\Big)\Gamma^{\wha\whb}-\tfrac{1}{72}r(W-\alpha\wedge\Psi)_{\wha\whb\whc}\Gamma^{\wha\whb\whc}\Big]\epsilon_\whpl
			\end{equation}
			and
			\begin{equation}\label{eq:KSEa-N>8Beta=0}
				\begin{aligned}
					0\ &=\ \Big[-\tfrac14\Big(r(\tilde\rmd\alpha)_{\wha\whb}+2r^2\dot\alpha_{[\wha}\alpha_{\whb]}\Big)\Gamma^\whb+\tfrac{1}{72}r(W-\alpha\wedge\Psi)_{\whb\whc\whd}\Gamma_\wha{}^{\whb\whc\whd}
					\\
					&\kern1cm-\tfrac{1}{12}r(W-\alpha\wedge\Psi)_{\wha\whb\whc}\Gamma^{\whb\whc}\Big]\epsilon_\whpl~,
				\end{aligned}
			\end{equation}
		\end{subequations}
		where we have used~\eqref{eq:uIndependenceOfSpinor}. Upon contracting~\eqref{eq:KSEa-N>8Beta=0} with $\Gamma_\wha$, we find
		\begin{equation}
			0\ =\ -\tfrac14\Big(r(\tilde\rmd\alpha)_{\wha\whb}+2r^2\dot\alpha_{[\wha}\alpha_{\whb]}\Big)\Gamma^{\wha\whb}\epsilon_\whpl~.
		\end{equation}
		Combining with~\eqref{eq:KSEu+N>8Beta=0}, we have
		\begin{equation}
			0\ =\ (W-\alpha\wedge\Psi)_{\wha\whb\whc}\Gamma^{\wha\whb\whc}~.
		\end{equation}
		With this,~\eqref{eq:KSEa-N>8Beta=0} can be rewritten as
		\begin{equation}\label{eq:WGamma=0N>8Int}
			0\ =\ \Big[-\tfrac14\Big(r(\tilde\rmd\alpha)_{\wha\whb}+2r^2\dot\alpha_{[\wha}\alpha_{\whb]}\Big)\Gamma^\whb-\tfrac18r(W-\alpha\wedge\Psi)_{\wha\whb\whc}\Gamma^{\whb\whc}\Big]\epsilon_\whpl~.
		\end{equation}
		The inner product of this with $\epsilon_\whpl$ is
		\begin{equation}
			0\ =\ \innerLarge{\epsilon_\whpl}{\big((\tilde\rmd\alpha)_{\wha\whb}+2r\dot\alpha_{[\wha}\alpha_{\whb]}\big)\Gamma^\whb\epsilon_\whpl}.
		\end{equation}
		One can use the same argument as in~\cite{Figueroa-OFarrill:2012kws} to conclude that $\innerLarge{\epsilon_\whpl}{\Gamma^\wha\epsilon_\whpl}$ span the space of tangent vectors on the spatial cross-section pointwise for $N>8$.\footnote{If $\inner{\epsilon_\whpl}{\Gamma_\wha\epsilon_\whpl}$ does not span the space of tangent vectors, there is a non-trivial vector $V$ such that $\inner{\epsilon_\whpl}{V_\wha\Gamma^\wha\epsilon_\whpl}=0$, Therefore, $V_\wha\Gamma^\wha\epsilon_\whpl$ is orthogonal to the space of Killing spinors. Since $N>8$, the vector space orthogonal to the space of Killing spinors has dimensions less than $8$. This implies that $V_\wha\Gamma^\wha$ has a non-trivial kernel. However, $V_\wha\Gamma^\wha V_\whb\Gamma^\whb=V^\wha V_\wha\id$. Hence, $V^\wha V_\wha=0$ and so, $V^\wha=0$ which contradicts the fact that $V$ is non-trivial.} Hence,
		\begin{equation}\label{eq:dalphaIdentityN>8}
			\tilde\rmd\alpha+r\dot\alpha\wedge\alpha\ =\ 0
		\end{equation}

		By multiplying~\eqref{eq:WGamma=0N>8Int} with $\Gamma_\whd$ and taking the inner product with $\epsilon_\whpl$, we obtain
		\begin{equation}
			0\ =\ r(W-\alpha\wedge\Psi)_{\wha\whc\whd}\innerLarge{\epsilon_\whpl\Gamma^\whc}{\epsilon_\whpl}.
		\end{equation}
		This implies that
		\begin{equation}\label{eq:WIdentityN>8}
			W-\alpha\wedge\Psi\ =\ 0~.
		\end{equation}

		\paragraph{Spinors in canonical form with $N=14$.}
		We first analyse the case of $N=14$ supersymmetry and show towards the end of the section that the proof also applies to $N=15,16$. For concreteness, we again view spinors as differential forms as summarised in \cref{App:SpinorForm}.
		
		Following the analysis in~\cite{Farotti:2021otm}, the space of $N=14$ positive chirality Killing spinors can be taken, without loss of generality, to be orthogonal with respect to $\inner{-}{-}$ to two normal spinors, $\nu^{(1)}_\whpl$ and $\nu^{(2)}_\whpl$, which can be simplified by applying appropriate $\sfSpin(9)$-transformations to
		\begin{equation}\label{eq:N=14beta=0TwoCanonicalSpinor}
			\nu^{(1)}_\whpl\ =\ 1+\bfe_{1234}
			\eand
			\nu^{(2)}_\whpl\ =\ \rmi q_1(1-\bfe_{1234})+q_2(\bfe_1+\bfe_{234})
		\end{equation}
		for $q_1,q_2\in\IR$. Consequently, the 14-dimensional space of Killing spinors is given by
		\begin{equation}\label{eq:caS14pl}
			\begin{aligned}
				\caS_{14}^\whpl\ &\coloneqq\ \spn_\IR\big\{\rmi q_2(1-\bfe_{1234})-q_1(\bfe_1+\bfe_{234}),\rmi(\bfe_1-\bfe_{234}),\bfe_{2}-\bfe_{134},\rmi(\bfe_2+\bfe_{134}),
				\\
				&\kern1cm\bfe_3+\bfe_{124},\rmi(\bfe_3-\bfe_{124}),\bfe_4-\bfe_{123},\rmi(\bfe_4+\bfe_{123}),\bfe_{12}-\bfe_{34},\rmi(\bfe_{12}+\bfe_{34}),\bfe_{13}+\bfe_{24},
				\\
				&\kern1cm\rmi(\bfe_{13}-\bfe_{24}),\bfe_{14}-\bfe_{23},\rmi(\bfe_{14}+\bfe_{23})\big\}\,.
			\end{aligned}
		\end{equation}
		The Killing spinor equation~\eqref{eq:KSEr-} now implies that $\Theta_\whpl\epsilon_\whpl=0$. Requiring that this condition hold for all $\epsilon_\whpl\in\caS_{14}^\whpl$ is sufficient to imply that $\Psi=0$. Hence, we can also conclude that $W=0$ as a consequence of~\eqref{eq:WIdentityN>8}. We also note that the gauge field equation~\eqref{eq:GaugeEoM:+-b} can be recast with the help of~\eqref{eq:dalphaIdentityN>8} as
		\begin{equation}\label{eq:FplusWedgeFplus}
			(X-r\alpha\wedge Z)\wedge(X-r\alpha\wedge Z)\ =\ 0~.
		\end{equation}
		We now divide the analysis into two cases.

		\paragraph{Case $q_1\neq 0$.}
		It is straightforward to show, component by component, that~\eqref{eq:FplusWedgeFplus} is not compatible with $\Theta_\whpl\epsilon_\whpl=0$ for all $\epsilon_\whpl\in\caS_{14}^\whpl$. The only solution that satisfies both conditions is
		\begin{equation}
			\alpha+r\dot\alpha\ =\ 0
			\eand
			X-r\alpha\wedge Z\ =\ 0~.
		\end{equation}
		In particular, the condition $\alpha+r\dot\alpha=0$ implies that
		\begin{equation}
			\caL_{\partial_r}(r\alpha)\ =\ 0
		\end{equation}
		and hence,
		\begin{equation}
			r\alpha\ =\ r\alpha\big|_{r=0}\ =\ 0
		\end{equation}
		from which it follows that $\alpha=0$ and hence, $X=0$ also.

		\paragraph{Case $q_1=0$, $q_2\neq 0$.}
		Unlike $q_1\neq 0$ case, there are non-trivial solutions, where both $\alpha$ and $X$ do not vanish and satisfy both~\eqref{eq:FplusWedgeFplus} and $\Theta_\whpl\epsilon_\whpl=0$ for all $\epsilon_\whpl\in\caS_{14}^\whpl$. Nonetheless, one can show by solving the linear system that such solutions satisfy
		\begin{equation}\label{eq:Lr(ralpha^2)AndF^2}
			(X-r\alpha\wedge Z)_{\wha\whb\whc\whd}(X-r\alpha\wedge Z)^{\wha\whb\whc\whd}\ =\ 1512 (\alpha_\wha+r\dot\alpha_\wha)(\alpha^\wha+r\dot\alpha^\wha)~.
		\end{equation}
		
		Recall the $\whpl\whmi$ component of the Einstein field equation~\eqref{eq:EinsteinEq+-} which now simplifies to
		\begin{equation}
			\begin{aligned}
				\tilde\nabla^\wha(\alpha_\wha+r\dot\alpha_\wha)\ &=\ \alpha_\wha\alpha^\wha-\tfrac{1}{72} (X-r\alpha\wedge Z)_{\wha\whb\whc\whd}(X-r\alpha\wedge Z)^{\wha\whb\whc\whd}
				\\
				&\kern1cm+r\Big(\tfrac12\alpha_\wha\alpha^\wha\dot\gamma_\whb{}^\whb-\alpha_\wha\alpha_\whb\dot\gamma^{\wha\whb}+4\alpha_\wha\dot\alpha^\wha\Big)
				\\
				&\kern1cm +r^2\Big(-\dot\alpha_\wha\alpha_\whb\dot\gamma^{\wha\whb}+\tfrac12\alpha^\wha\dot\alpha_\wha\dot\gamma_\whb{}^\whb+\dot\alpha_\wha\dot\alpha^\wha+\alpha_\wha\ddot\alpha^\wha\Big)\,
			\end{aligned}
		\end{equation} 
		where we have used $\beta=0$, $\Psi=0$, and~\eqref{eq:WIdentityN>8}. This simplifies further to
		\begin{equation}
			\begin{aligned}
				\tilde\nabla^\wha(\alpha_\wha+r\dot\alpha_\wha)\ &=\ \caL_{\partial_r}[r\alpha^\wha(\alpha_\wha+r\dot\alpha_\wha)]+\tfrac12\dot\gamma_\whb{}^\whb r\alpha^\wha(\alpha_\wha+r\dot\alpha_\wha)
				\\
				&\kern1cm-\tfrac{1}{72}(X-r\alpha\wedge Z)_{\wha\whb\whc\whd}(X-r\alpha\wedge Z)^{\wha\whb\whc\whd}~.
			\end{aligned}
		\end{equation}
		Integration over the spatial cross-section $S$ yields
		\begin{equation}\label{eq:IntDivergenceOfAlpha}
			\tfrac{1}{72}\int_S(X-r\alpha\wedge Z)_{\wha\whb\whc\whd}(X-r\alpha\wedge Z)^{\wha\whb\whc\whd}\ =\ \caL_{\partial_r}\int_S(r\alpha^\wha\alpha_\wha+r^2\alpha^\wha\dot\alpha_\wha)~.
		\end{equation}
		
		We will now prove by induction that $\caL_{\partial_r}^{(n)}\alpha\big|_{r=0}=0$ and $\caL_{\partial_r}^{(n)}X\big|_{r=0}=0$ for all $n\geq0$ where $\caL_{\partial_r}^{(0)}\alpha\big|_{r=0}\coloneqq\alpha|_{r=0}=\mathring\alpha$ and $\caL_{\partial_r}^{(0)}X\big|_{r=0}\coloneqq X|_{r=0}=\mathring X$. This then implies that $\alpha=0$ and $X=0$.
		
		To prove the case $n=0$, we evaluate both~\eqref{eq:Lr(ralpha^2)AndF^2} and~\eqref{eq:IntDivergenceOfAlpha} at $r=0$ which yield
		\begin{equation}
			\mathring X_{\wha\whb\whc\whd}\mathring X^{\wha\whb\whc\whd}\ =\ 1512\mathring\alpha_\wha\mathring\alpha^\wha
			\eand
			\tfrac{1}{72}\int_{\mathring S}\mathring X_{\wha\whb\whc\whd}\mathring X^{\wha\whb\whc\whd}\ =\ \int_{\mathring S}\mathring\alpha_\wha\mathring\alpha^\wha~.
		\end{equation}
		These imply that $\mathring X_{\wha\whb\whc\whd}\mathring X^{\wha\whb\whc\whd}=\mathring\alpha^\wha\mathring\alpha_\wha=0$. Hence, $\mathring\alpha=0$ and $\mathring X=0$.
		
		We now assume that $\caL_{\partial_r}^{(n)}\alpha\big|_{r=0}=0$ and $\caL_{\partial_r}^{(n)}X\big|_{r=0}=0$ for all $0\leq n\leq m$ where $m\in\IN$. To show that $\caL_{\partial_r}^{(m+1)}\alpha\big|_{r=0}=0$ and $\caL_{\partial_r}^{(m+1)}X\big|_{r=0}=0$, we apply $\caL_{\partial_r}^{(2m+2)}$ on both~\eqref{eq:Lr(ralpha^2)AndF^2} and~\eqref{eq:IntDivergenceOfAlpha} and evaluate at $r=0$. By our induction hypothesis, we obtain
		\begin{subequations}
			\begin{equation}
				\big(\caL_{\partial_r}^{(m+1)}X\big)_{\wha\whb\whc\whd}\big(\caL_{\partial_r}^{(m+1)}X\big)^{\wha\whb\whc\whd}\Big|_{r=0}\ =\ 1512(m+2)^2\big(\caL_{\partial_r}^{(m+1)}\alpha\big)_\wha\big(\caL_{\partial_r}^{(m+1)}\alpha\big)^\wha\Big|_{r=0}
			\end{equation}
			and
			\begin{equation}
				\tfrac{1}{72}\int_{\mathring S}\big(\caL_{\partial_r}^{(m+1)}X\big)_{\wha\whb\whc\whd}\big(\caL_{\partial_r}^{(m+1)}X\big)^{\wha\whb\whc\whd}\Big|_{r=0}\ =\ (2m^2+7m+6)\int_{\mathring S}\big(\caL_{\partial_r}^{(m+1)}\alpha\big)_\wha\big(\caL_{\partial_r}^{(m+1)}\alpha\big)^\wha\Big|_{r=0}~.
			\end{equation}
		\end{subequations}
		Upon combining these two equations, we obtain
		\begin{equation}
			(78+77m+19m^2)\int_{\mathring S}\big(\caL_{\partial_r}^{(m+1)}\alpha\big)_\wha\big(\caL_{\partial_r}^{(m+1)}\alpha\big)^\wha\Big|_{r=0}\ =\ 0~.
		\end{equation}
		Since $78+77m+19m^2>0$ for all $m\geq 0$, we conclude that $\caL_{\partial_r}^{(m+1)}\alpha\big|_{r=0}=0$ and $\caL_{\partial_r}^{(m+1)}X\big|_{r=0}=0$. Therefore, by induction, we have proved that $\caL_{\partial_r}^{(n)}\alpha\big|_{r=0}=0$ and $\caL_{\partial_r}^{(n)} X\big|_{r=0}=0$ for all $n\geq0$.
		
		\paragraph{pp-wave solutions.}
		Having established that $\alpha=0$, $X=0$, $\Psi=0$, and $W=0$ for any non-trivial $\nu^{(2)}_\whpl$ defined in~\eqref{eq:N=14beta=0TwoCanonicalSpinor}, we show now that the solutions are pp-waves. Firstly, we consider the Killing spinor equation~\eqref{eq:KSEa+} which implies that
		\begin{equation}
			\tilde\nabla_\wha\epsilon_\whpl\ =\ 0
		\end{equation}
		and so,
		\begin{equation}
			\tilde R_{\wha\whb\whc\whd}\Gamma^{\whc\whd}\epsilon_\whpl\ =\ 0
		\end{equation}
		for any $\epsilon_\whpl\in\caS_{14}^\whpl$. This is sufficient to imply that the Riemann tensor $\tilde R_{\wha\whb\whc\whd}$ vanishes.

		Next, the Bianchi identities~\eqref{eq:GaugeBianchi} imply that
		$\tilde\rmd Z=0$, and the gauge equation~\eqref{eq:GaugeEoM:-b1b2}
		imply that $\tilde\nabla^\wha Z_{\wha\whb\whc}=0$. These conditions, together with $\tilde R_{\wha\whb\whc\whd}=0$, yield $\tilde\nabla_\wha Z_{\whb\whc\whd}=0$. We then return to the integrability condition~\eqref{eq:ra+Integrability} which simplifies to
		\begin{equation}
			\tilde\nabla_{[\wha}\dot\gamma_{\whb]\whc}\Gamma^{\wha\whb}\epsilon_\whpl\ =\ 0
		\end{equation}
		for any $\epsilon_\whpl\in\caS_{14}$. Hence,
		\begin{equation}\label{eq:GammaDotCodazzi}
			\tilde\nabla_{[\wha}\dot\gamma_{\whb]\whc}\ =\ 0
		\end{equation}
		and so, $\dot\gamma$ is a Codazzi tensor.

		To proceed, we take a further Lie derivative of~\eqref{eq:GammaDotCodazzi} with respect to $\partial_r$. We find
		\begin{subequations}
			\begin{equation}
				\tilde\nabla_{[\wha}M_{\whb]\whc}\ =\ 0
			\end{equation}
			with
			\begin{equation}
				M_{\wha\whb}\ \coloneqq\ \ddot\gamma_{\wha\whb}-\tfrac12\dot\gamma^\whc{}_\wha\dot\gamma_{\whc\whb}~.
			\end{equation}
		\end{subequations}
		We observe that also $M$ is a Codazzi tensor with the trace given by
		\begin{equation}
			\gamma^{\wha\whb}M_{\wha\whb}\ =\ -\tfrac16Z_{\wha\whb\whc}Z^{\wha\whb\whc}
		\end{equation}
		as a consequence of the Einstein field equation~\eqref{eq:EinsteinEq--}. Furthermore, the condition $\tilde\nabla_\wha Z_{\whb\whc\whd}=0$ implies that $\gamma^{\wha\whb}M_{\wha\whb}$ depends only on $r$. Consequently, it follows by a compactness argument that $M$ is parallel on the spatial cross-section; $\tilde\nabla_\wha M_{\whb\whc}=0$ (see~\cite[Theorem 16.9]{besse2007einstein} and references therein). In turn, the condition $\tilde\nabla_\wha M_{\whb\whc}=0$ is equivalent to
		\begin{equation}\label{eq:covdercod}
			\caL_{\partial_r}\tilde\nabla_\wha\dot\gamma_{\whb\whc}\ =\ 0~.
		\end{equation}
		
		We can, without loss of generality, choose to work in a basis for which $\dot\gamma_\wha{}^\wha\big|_{r=0}$ is constant~\cite{Gutowski:2025lzi}. It follows from a compactness argument that $\tilde\nabla_\wha\dot\gamma_{\whb\whc}\big|_{r=0}=0$. Then, the condition~\eqref{eq:covdercod} implies that $\tilde\nabla_\wha\dot\gamma_{\whb\whc}=0$.

		It will be convenient to work with local coordinates $y^i$ on the spatial cross-section. As $\tilde\nabla_\wha\dot\gamma_{\whb\whc}=0$, this implies that $\partial_r\omega^i_{jk}=0$, so $\omega^i_{jk}={\mathring\omega}{}^i_{jk}$, where $\omega$ and $\mathring\omega$ are Christoffel symbols with respect to $\gamma$ and $\mathring\gamma$, respectively. As $\mathring\gamma$ is flat (recall that $\tilde R_{\wha\whb\whc\whd}=0$), we can make an $r$-independent coordinate transformation (which preserves the form of the metric in Gau{\ss}ian null coordinates) in order to set, without loss of generality, $\mathring\gamma_{ij}=\delta_{ij}$. Moreover, it then follows that $\Gamma^i_{jk}=\mathring\Gamma{}^i_{jk}=0$ and hence $\partial_i\gamma_{jk}=0$, so $\gamma_{ij}=\gamma_{ij}(r)$ with $\gamma_{ij}(0)=\delta_{ij}$. We remark that the resulting metric
		\begin{equation}\label{eq:ppwavesol}
			g\ =\ 2\rmd u\odot\rmd r+\tfrac12\gamma_{ij}(r)\rmd y^i\odot\rmd y^j
		\end{equation}
		is that of a plane wave written in Rosen coordinates. The coordinate transformation which can be used to write this metric in Brinkman coordinates can be found in~\cite[Section 2.2]{Blau:2002mw}, and this transformation preserves the form of the four-form $F$. It is known that such geometries do not correspond to black holes as they do not have an event horizon~\cite{Hubeny:2002pj,Hubeny:2003ug}.
		
		\paragraph{$N=15,16$ supersymmetry.}
		For the case of $N=16$ supersymmetry, the space of Killing spinors is  the space of all positive chirality Killing spinors, i.e.
		\begin{equation}
			\caS_{16}^\whpl\ =\ \spn_\IR\big\{\nu_\whpl^{(1)}\big\}\oplus\spn_\IR\big\{\nu_\whpl^{(2)}\big\}\oplus\caS_{14}^\whpl~,
		\end{equation}
		where $\nu_\whpl^{(1)}$, $\nu_\whpl^{(2)}$, and $\caS_{14}^\whpl$ are defined in~\eqref{eq:N=14beta=0TwoCanonicalSpinor} and~\eqref{eq:caS14pl}.
		
		For the case of $N=15$ supersymmetry, the space of Killing spinors can be taken, without loss of generality, to be orthogonal to the spinor 
		\begin{equation}
			\nu^{(1)}_\whpl\ =\ 1+\bfe_{1234}~,
		\end{equation}
		which is identical to $\nu^{(1)}_\whpl$ in~\eqref{eq:N=14beta=0TwoCanonicalSpinor}. Hence, the 15-dimensional space of Killing spinors is
		\begin{equation}
			\caS_{15}^\whpl\ =\ \spn_\IR\big\{\nu_\whpl^{(2)}\big\}\oplus\caS_{14}^\whpl~,
		\end{equation}
		where $\nu_\whpl^{(2)}$ and $\caS_{14}^\whpl$ are defined in~\eqref{eq:N=14beta=0TwoCanonicalSpinor} and~\eqref{eq:caS14pl}.

		Since we have the vector space inclusions
		\begin{equation}
			\caS_{14}^\whpl\ \subseteq\ \caS_{15}^\whpl\ \subseteq\ \caS_{16}^\whpl~,
		\end{equation}
		the results for $N=14$ also hold when the spaces of the Killing spinors are $\caS_{15}^\whpl$ and $\caS_{16}^\whpl$. In conclusion, for $N\geq14$ supersymmetry with $\beta=0$, the solutions are pp-waves given by~\eqref{eq:ppwavesol}.

		\appendix
		\addappheadtotoc
		\appendixpage

		\appendices

		\section{Gauge field equation}\label{App:GaugeFieldEqs}

		By decomposing $F$ as in~\eqref{eq:GaugeFieldInOrthonormal}, the gauge field equation~\eqref{eq:GaugeEqsShort} reads explicitly as follows.
		\begin{subequations}
			\begin{itemize}
				\item $\whpl\whmi\wha$ components:
					\begin{equation}\label{eq:GaugeEoM:+-b}
						\begin{aligned}
							0\ &=\ \tilde\nabla^\whb\Psi_{\whb\wha}+r\Big[-\alpha^\whb\dot\Psi_{\whb\wha}+\alpha^\whb\dot\gamma_\whb{}^\whc\Psi_{\whc\wha}-\tfrac12\alpha^\whc\dot\gamma_\whb{}^\whb\Psi_{\whc\wha}+\alpha^\whb\dot\gamma^\whc{}_\wha\Psi_{\whb\whc}
							\\
							&\kern1cm-\tfrac12(\tilde\rmd\alpha)^{\whb\whc}Z_{\whb\whc\wha}\Big]-r^2\dot\alpha^\whb\alpha^\whc Z_{\whb\whc\wha}
							\\
							&\kern1cm+\tfrac{1}{2(4!)^2}\epsilon_{\whpl\whmi\wha}{}^{\wha_1\cdots\wha_8}\Big[X_{\wha_1\cdots\wha_4}X_{\wha_5\cdots\wha_8}-2rX_{\wha_1\cdots\wha_4}(\alpha\wedge Z)_{\wha_5\cdots\wha_8}\Big]\,,
						\end{aligned}
					\end{equation}
				\item $\whpl\wha\whb$ components:
					\begin{equation}
						\begin{aligned}
							0\ &=\ r\Big[-\tilde\nabla^\whc(W-\alpha\wedge\Psi)_{\whc\wha\whb}+\alpha^\whc(W-\alpha\wedge\Psi)_{\whc\wha\whb}-\tfrac12(\tilde\rmd\alpha)^{\whc\whd}X_{\whc\whd\wha\whb}\Big]
							\\
							&\kern1cm
							+r^2\Big[-\tfrac12\beta\dot\Psi_{\wha\whb}+\alpha^\whc(\dot W-\dot\alpha\wedge\Psi-\alpha\wedge\dot\Psi)_{\whc\wha\whb}
							\\
							&\kern1cm+(-\alpha^\whc\dot\gamma^\whd{}_\whc+\tfrac12\alpha^\whd\dot\gamma^\whe{}_\whe)(W-\alpha\wedge\Psi)_{\whd\wha\whb}-2\alpha^\whd\dot\gamma^\whc{}_{[\wha|}(W-\alpha\wedge\Psi)_{\whd\whc|\whb]}
							\\
							&\kern1cm+\beta\dot\gamma^\whc{}_{[\wha|}\Psi_{\whc|\whb]}-\tfrac14\beta\dot\gamma^\whc{}_\whc\Psi_{\wha\whb}-\tfrac12\beta\tilde\nabla^\whc Z_{\whc\wha\whb}-(\tilde\rmd\beta)^\whc Z_{\whc\wha\whb}
							\\
							&\kern1cm+\tfrac32\alpha^\whc\beta Z_{\whc\wha\whb}-\tfrac12(\tilde\rmd\alpha)^{\whc\whd}(\alpha\wedge Z)_{\whd\whc\wha\whb}-\dot\alpha^\whc\alpha^\whd X_{\whc\whd\wha\whb}\big]
							\\
							&\kern1cm+r^3\big[\dot\beta\alpha^\whc Z_{\whc\wha\whb}+\tfrac12\beta\alpha^\whc\dot Z_{\whc\wha\whb}-\tfrac12\beta\dot\alpha^\whc Z_{\whc\wha\whb}
							\\
							&\kern1cm+\tfrac12\beta Z_{\whc\wha\whb}(-\alpha^\whd\dot\gamma^\whc{}_\whd+\tfrac12\alpha^\whc\dot\gamma^\whd{}_\whd)-\beta\alpha^\whc\dot\gamma^\whd{}_{[\wha|}Z_{\whc\whd|\whb]}+\dot\alpha^\whc\alpha^\whd(\alpha\wedge Z)_{\whc\whd\wha\whb}\big]
							\\
							&\kern1cm+\epsilon_{\whpl\whmi\wha\whb}{}^{\wha_1\ldots\wha_7}\big[\tfrac{r}{144}(W-\alpha\wedge\Psi)_{\wha_1\wha_2\wha_3}X_{\wha_4\wha_5\wha_6\wha_7}-\tfrac{r^2}{36}W_{\wha_1\wha_2\wha_3}\alpha_{\wha_4}Z_{\wha_5\wha_6\wha_7}
							\\
							&\kern1cm+\tfrac{r^2}{288}\beta Z_{\wha_1\wha_2\wha_3}X_{\wha_4\wha_5\wha_6\wha_7}\big]\,,
						\end{aligned}
					\end{equation}
				\item $\whmi\wha\whb$ components:
					\begin{equation}\label{eq:GaugeEoM:-b1b2}
						\begin{aligned}
							0\ &=\ \dot\Psi_{\wha\whb}+\tfrac12\dot\gamma^\whc{}_\whc\Psi_{\wha\whb}-\tilde\nabla^\whc Z_{\whc\wha\whb}-2\dot\gamma^\whc{}_{[\wha|}\Psi_{\whc|\whb]}+\alpha^\whc Z_{\whc\wha\whb}
							\\
							&\kern1cm+r\big[\alpha^\whc\dot Z_{\whc\wha\whb}+\dot\alpha^\whc Z_{\whc\wha\whb}+\tfrac12\alpha^\whc\dot\gamma^\whd{}_\whd Z_{\whc\wha\whb}-\alpha^\whd\dot\gamma^\whc{}_\whd Z_{\whc\wha\whb}
							\\
							&\kern1cm+2\alpha^\whd\dot\gamma^\whc{}_{[\wha|}Z_{\whc\whd|\wha]}\big]
							\\
							&\kern1cm-\tfrac{1}{144}\epsilon_{\whpl\whmi\wha\whb}{}^{\wha_1\ldots\wha_7}Z_{\wha_1\wha_2\wha_3}X_{\wha_4\wha_5\wha_6\wha_7}~,
						\end{aligned}
					\end{equation}
				\item $\wha\whb\whc$ components:
					\begin{equation}
						\begin{aligned}
							0\ &=\ \tilde\nabla^\whd X_{\whd\wha\whb\whc}+(W-\alpha\wedge\Psi)_{\wha\whb\whc}-\alpha^\whd X_{\whd\wha\whb\whc}
							\\
							&\kern1cm+r\big[-\tilde\nabla^\whd(\alpha\wedge Z)_{\whd\wha\whb\whc}+\dot W_{\wha\whb\whc}-(\dot\alpha\wedge\Psi)_{\wha\whb\whc}-(\alpha\wedge\dot\Psi)_{\wha\whb\whc}
							\\
							&\kern1cm+2\beta Z_{\wha\whb\whc}-\dot\alpha^\whd X_{\whd\wha\whb\whc}-\alpha^\whd(\dot X-2\alpha\wedge Z)_{\whd\wha\whb\whc}
							\\
							&\kern1cm+\tfrac12\dot\gamma^\whd{}_\whd(W-\alpha\wedge\Psi)_{\wha\whb\whc}-3\dot\gamma^\whd{}_{[\wha|}(W-\alpha\wedge\Psi)_{\whd|\whb\whc]}+\alpha^\whd\dot\gamma^\whe{}_\whd X_{\whe\wha\whb\whc}
							\\
							&\kern1cm-\tfrac12\alpha^\whd\dot\gamma^\whe{}_\whe X_{\whd\wha\whb\whc}+3\alpha^\whd\dot\gamma^\whe{}_{[\wha|}X_{\whd\whe|\whb\whc]}\big]
							\\
							&\kern1cm+r^2\big[\dot\beta Z_{\wha\whb\whc}+\beta\dot Z_{\wha\whb\whc}+\alpha^\whd(\dot\alpha\wedge Z+\alpha\wedge\dot Z)_{\whd\wha\whb\whc}+\dot\alpha^\whd(\alpha\wedge Z)_{\whd\wha\whb\whc}
							\\
							&\kern1cm-3\beta\dot\gamma^\whd{}_{[\wha|}Z_{\whd|\whb\whc]}+\tfrac12\dot\gamma^\whd{}_\whd\beta Z_{\wha\whb\whc}-\alpha^\whd\dot\gamma^\whe{}_\whd(\alpha\wedge Z)_{\whe\wha\whb\whc}
							\\
							&\kern1cm+\tfrac12\alpha^\whd\dot\gamma^\whe{}_\whe(\alpha\wedge Z)_{\whd\wha\whb\whc}-3\alpha^\whd\dot\gamma^\whe{}_{[\wha|}(\alpha\wedge Z)_{\whd\whe|\whb\whc]}\big]
							\\
							&\kern1cm+\epsilon_{\whpl\whmi\wha\whb\whc}{}^{\wha_1\ldots\wha_6}\big[\tfrac{r}{36}(W-\alpha\wedge\Psi)_{\wha_1\wha_2\wha_3}Z_{\wha_4\wha_5\wha_6}
							\\
							&\kern1cm-\tfrac{1}{48}\Psi_{\wha_1\wha_2}(X-r\alpha\wedge Z)_{\wha_3\wha_4\wha_5\wha_6}\big]\,,
						\end{aligned}
					\end{equation}
			\end{itemize}
		\end{subequations}
		where $\tilde\nabla$ is the Levi-Civita connection with respect to the metric $\gamma$ on the spatial cross-section $S$.

		\section{Connection one-form}\label{App:Connection1Form}

		In the null orthonormal basis~\eqref{eq:OrthonormalBasis}, the components of connection one-form are given by
		\begin{equation}
			\tilde\omega_{\whA\whB}{}^\whC\ =\ \tfrac12(C^\whC{}_{\whA\whB}+C^\whC{}_{\whB\whA}+C_{\whA\whB}{}^\whC)
			\ewith
			\rmd e^\whA\ =\ \tfrac12e^\whC\wedge e^\whB C_{\whB\whC}{}^\whA~.
		\end{equation}
		Explicitly, we have
		\begin{equation}
			\begin{aligned}
				\omega_\whmi{}^\whmi\ &=\ -\omega_\whpl{}^\whpl
				\\
				&=\ -e^\whpl r\big(\beta+\tfrac12r\dot\beta\big)+e^\wha\tfrac12\big(\alpha_\wha+r\dot\alpha_\wha\big)\,,
				\\
				\omega_\whpl{}^\wha\ &=\ -\omega_\wha{}^\whmi
				\\
				&=\ -e^\whpl\tfrac12r^2\big[\beta\alpha_\wha-\tilde\nabla_\wha\beta-r\big(\beta\dot\alpha_\wha-\dot\beta\alpha_\wha\big)\big]
				\\
				&\kern1cm+e^\whmi\tfrac12\big(\alpha_\wha+r\dot\alpha_\wha\big)
				\\
				&\kern1cm-e^\whb\tfrac12r\big[(\tilde{\rmd}\alpha)_{\wha\whb}+r\big(\dot\alpha_\wha\alpha_\whb-\dot\alpha_\whb\alpha_\wha\big)-\tfrac12r\beta\big(\dot e_\whb{}^\wha+\dot e_\whc{}^\whd\delta^{\whc\wha}\delta_{\whd\whb}\big)\big]\,,
				\\
				\omega_\whmi{}^\wha\ &=\ -\omega_\wha{}^\whpl
				\\
				&=\ e^\whpl\tfrac12\big(\alpha_\wha+r\dot\alpha_\wha\big)+e^\whb\tfrac12\big(\dot e_\whb{}^\wha+\dot e_\whc{}^\whd\delta^{\whc\wha}\delta_{\whd\whb}\big)\,,
				\\
				\omega_\whb{}^\wha\ &=\ e^\whpl\tfrac12r\big[(\tilde{\rmd}\alpha)_\whb{}^\wha-r\big(\alpha_\whb\dot\alpha^\wha-\dot\alpha_\whb\alpha^\wha\big)-\tfrac12r\beta\big(\dot e_\whb{}^\wha-\dot e_\whc{}^\whd\delta^{\whc\wha}\delta_{\whd\whb}\big)\big]
				\\
				&\kern1cm-\tfrac12e^\whmi\big(\dot e_\whb{}^\wha-\dot e_\whc{}^\whd\delta^{\whc\wha}\delta_{\whd\whb}\big)
				\\
				&\kern1cm+e^\whc\big[r\big(\alpha_{[\whc}\dot e_{\whb]}{}^\wha+\alpha_{[\whd}\dot e_{\whc]}{}^\whe\delta^{\whd\wha}\delta_{\whe\whb}+\alpha_{[\whd}\dot e_{\whb]}{}^\whe\delta^{\whd\wha}\delta_{\whe\whc}\big)+\tilde\omega_{\whc\whb}{}^\wha\big]\,.
			\end{aligned}
		\end{equation}
		Note that $\dot e_\whb{}^\wha$ are the $\whb$-components in the frame basis~\eqref{eq:OrthonormalBasis} of the Lie derivative of the one-form $e^\wha$ with respect to $\partial_r$, i.e.
		\begin{equation}\label{eq:Defedot}
			\dot e_\whb{}^\wha\ \coloneqq\ E_\whb{}^i\partial_r(e_i{}^\wha)~.
		\end{equation}
		Furthermore, $\tilde\omega_{\wha\whb}{}^\whc$ are the components of the connection one-form on the spatial cross-section $S$,
		\begin{subequations}
			\begin{equation}
				\tilde\omega_{\wha\whb}{}^\whc\ =\ \tfrac12(\tilde C^\whc{}_{\wha\whb}+\tilde C^\whc{}_{\whb\wha}+\tilde C_{\wha\whb}{}^\whc)
			\end{equation}
			with
			\begin{equation}
				\tilde\rmd e^\wha\ =\ \tfrac12e^\whc\wedge e^\whb\tilde C_{\whb\whc}{}^\wha~,
			\end{equation}
		\end{subequations}
		and the indices are raised and lowered by $\delta^{\wha\whb}$ and $\delta_{\wha\whb}$ respectively.

		\section{Einstein field equation}\label{App:EinsteinEqs}

		The Einstein field equation is
		\begin{equation}\label{eq:EinsteinEqs}
			R_{\whA\whB}\ =\ \tfrac{1}{12}F_\whA{}^{\whC\whD\whE}F_{\whB\whC\whD\whE}-\tfrac{1}{144}g_{\whA\whB}F_{\whC\whD\whE\whF}F^{\whC\whD\whE\whF}~.
		\end{equation}
		By using~\eqref{eq:OrthonormalBasis} and~\eqref{eq:GaugeFieldInOrthonormal}, we have for the right-hand side
		\begin{subequations}\label{eq: EinsteinEq}
			\begin{equation}\label{eq:EinsteinEq--}
				R_{\whmi\whmi}\ =\ \tfrac{1}{12}Z^2
			\end{equation}
			and
			\begin{equation}\label{eq:EinsteinEq+-}
				\begin{aligned}
					R_{\whpl\whmi}\ &=\ -\tfrac{1}{144}X_{\wha\whb\whc\whd}X^{\wha\whb\whc\whd}-\tfrac16\Psi_{\wha\whb}\Psi^{\wha\whb}
					\\
					&\kern1cm +r\Big[\tfrac{1}{36}\big(W-\alpha\wedge\Psi+\tfrac12 r\beta Z\big)_{\wha\whb\whc}Z^{\wha\whb\whc}+\tfrac{1}{72}(\alpha\wedge Z)_{\wha\whb\whc\whd}X^{\wha\whb\whc\whd}\Big]
					\\
					&\kern1cm-\tfrac{1}{144}r^2(\alpha\wedge Z)_{\wha\whb\whc\whd}(\alpha\wedge Z)^{\wha\whb\whc\whd}
				\end{aligned}
			\end{equation}
			and
			\begin{equation}
				R_{\whpl\whpl}\ =\ \tfrac{r^2}{12}(W-\alpha\wedge\Psi)_{\wha\whb\whc}(W-\alpha\wedge\Psi)^{\wha\whb\whc}+\tfrac{r^3}{12}\beta(W-\alpha\wedge\Psi)_{\wha\whb\whc}Z^{\wha\whb\whc}+\tfrac{r^4}{48}\beta^2Z_{\wha\whb\whc}Z^{\wha\whb\whc}
			\end{equation}
			and
			\begin{equation}\label{eq:EinsteinEq-b}
				R_{\whmi\wha}\ =\ \tfrac14Z_{\wha\whb\whc}\Psi^{\whb\whc}+\tfrac{1}{12}X_{\wha\whb\whc\whd}Z^{\whb\whc\whd}-\tfrac{r}{12}(\alpha\wedge Z)_{\wha\whb\whc\whd}Z^{\whb\whc\whd}
			\end{equation}
			and
			\begin{equation}\label{eq: EinsteinEq+b}
				\begin{aligned}
					R_{\whpl\wha}\ &=\ r\Big[-\tfrac14(W-\alpha\wedge\Psi)_{\wha\whb\whc}\Psi^{\whb\whc}+\tfrac{1}{12}X_{\wha\whb\whc\whd}(W-\alpha\wedge\Psi)^{\whb\whc\whd}\Big]
					\\
					&\kern1cm+r^2\Big[-\tfrac18\beta Z_{\wha\whb\whc}\Psi^{\whb\whc}-\tfrac{1}{12}(\alpha\wedge Z)_{\wha\whb\whc\whd}(W-\alpha\wedge\Psi)^{\whb\whc\whd}
					\\
					&\kern1cm+\tfrac{1}{24}\beta X_{\wha\whb\whc\whd}Z^{\whb\whc\whd}\Big]-\tfrac{r^3}{24}\beta(\alpha\wedge Z)_{\wha\whb\whc\whd}Z^{\whb\whc\whd}
				\end{aligned}
			\end{equation}
			and
			\begin{equation}
				\begin{aligned}
					R_{\wha\whb}\ &=\ -\tfrac12\Psi_{\wha\whc}\Psi_\whb{}^\whc+\tfrac{1}{12}X_{\wha\whc\whd\whe}X_\whb{}^{\whc\whd\whe}+\delta_{\wha\whb}\Big(\tfrac{1}{12}\Psi_{\whc\whd}\Psi^{\whc\whd}-\tfrac{1}{144}X_{\whc\whd\whe\whf}X^{\whc\whd\whe\whf}\Big)
					\\
					&\kern1cm+r\Big\{\tfrac12(W-\alpha\wedge\Psi)_{(\wha|\whc\whd}Z_{|\whb)}{}^{\whc\whd}-\tfrac16(\alpha\wedge Z)_{(\wha|\whc\whd\whe}X_{|\whb)}{}^{\whc\whd\whe}
					\\
					&\kern1cm+\delta_{\wha\whb}\Big[\tfrac{1}{72}(\alpha\wedge Z)_{\whc\whd\whe\whf}X^{\whc\whd\whe\whf}-\tfrac{1}{18}Z_{\whc\whd\whe}(W-\alpha\wedge\Psi)^{\whc\whd\whe}\Big]\Big\}
					\\
					&\kern1cm +r^2\Big\{\tfrac{1}{12}(\alpha\wedge Z)_{\wha\whc\whd\whe}(\alpha\wedge Z)_\whb{}^{\whc\whd\whe}+\tfrac14\beta Z_{\wha\whc\whd}Z_{\whb}{}^{\whc\whd}
					\\
					&\kern1cm+\delta_{\wha\whb}\Big[-\tfrac{1}{36}\beta Z_{\whc\whd\whe}Z^{\whc\whd\whe}-\tfrac{1}{144}(\alpha\wedge Z)_{\whc\whd\whe\whf}(\alpha\wedge Z)^{\whc\whd\whe\whf}\Big]\Big\}\,.
				\end{aligned}
			\end{equation}
		\end{subequations}
		Furthermore, the components of the Ricci tensor on the left-hand side of~\eqref{eq:EinsteinEqs} are given by
		\begin{subequations}
			\begin{equation}
				\begin{aligned}
					R_{\whmi\whmi}\ &=\ -\tfrac12\ddot\gamma_\wha{}^\wha+\tfrac14\dot\gamma^{\wha\whb}\dot\gamma_{\wha\whb}
				\end{aligned}
			\end{equation}
			and
			\begin{equation}
				\begin{aligned}
					R_{\whpl\whmi}\ &=\ -\beta+\tfrac12\tilde\nabla^\wha\alpha_\wha-\tfrac12\alpha^\wha\alpha_\wha
					\\
					&\kern1cm+r\Big(-2\dot\beta-\tfrac12\beta\dot\gamma_\wha{}^\wha-2\dot\alpha^\wha\alpha_\wha+\tfrac12\alpha_\wha\alpha_\whb\dot\gamma^{\wha\whb}-\tfrac14\alpha_\wha\alpha^\wha\dot\gamma_\whb{}^\whb+\tfrac12\tilde\nabla^\wha\dot\alpha_\wha\Big)
					\\
					&\kern1cm+r^2\Big[-\tfrac12\ddot\beta-\tfrac12\ddot\alpha^\wha\alpha_\wha-\tfrac14\beta\ddot\gamma_\wha{}^\wha-\tfrac14\Big(\dot\beta+\dot\alpha^\whb\alpha_\whb\Big){\dot\gamma}_\wha{}^\wha+\tfrac12\dot\gamma^{\wha\whb}\dot\alpha_\wha\alpha_\whb
					\\
					&\kern1cm-\tfrac12\dot\alpha^\wha\dot\alpha_\wha+\tfrac18\beta\dot\gamma^{\wha\whb}\dot\gamma_{\wha\whb}\Big]
				\end{aligned}
			\end{equation}
			and
			\begin{equation}
				\begin{aligned}
					R_{\whpl\whpl}\ &=\ r^2\Big\{\beta\alpha_\wha\alpha^\wha-\tfrac32\alpha^\wha \tilde\nabla_\wha\beta+\tilde\nabla_{[\wha }\alpha_{\whb ]}\tilde\nabla^\wha\alpha^\whb-\tfrac12\beta\tilde\nabla^\wha\alpha_\wha+\tfrac12\tilde\nabla^\wha\tilde\nabla_\wha\beta\Big\}
					\\
					&\kern1cm+r^3\Big[\tfrac14\dot\gamma_\wha{}^\wha\left(\beta\alpha_\whb\alpha^\whb-\alpha_\whb\tilde\nabla^\whb\beta\right)+\tfrac12\dot\gamma^{\wha\whb}\left(-\beta\alpha_\wha\alpha_\whb+\alpha_\wha\tilde\nabla_\whb\beta\right)
					\\
					&\kern1cm+\tfrac12\beta\tilde\nabla^\wha\dot\alpha_\wha-\tfrac12\dot\beta\tilde\nabla^\wha\alpha_\wha+2\dot\beta\alpha^\wha\alpha_\wha-\alpha^\wha\tilde\nabla_\wha\dot\beta+\tfrac12\dot\alpha^\wha\tilde\nabla_\wha\beta-\beta\dot\alpha^\wha\alpha_\wha
					\\
					&\kern1cm-2\dot\alpha^\wha\alpha^\whb\tilde\nabla_{[\whb}\alpha_{\wha]}\Big]
					\\
					&\kern1cm+r^4\Big[\tfrac14\dot\gamma_\wha{}^\wha\Big(-\beta\dot\alpha_\whb\alpha^\whb+\dot\beta\alpha_\whb\alpha^\whb\Big)+\tfrac12\dot\gamma^{\wha\whb}\Big(\beta\dot\alpha_\wha\alpha_\whb-\dot\beta\alpha_\wha\alpha_\whb\Big)
					\\
					&\kern1cm-\tfrac18\beta^2\ddot\gamma_\wha{}^\wha-\tfrac12\beta\ddot\alpha^\wha\alpha_\wha+\tfrac12\ddot\beta{\alpha}^\wha\alpha_\wha+\tfrac12\dot\alpha^\wha\alpha^\whb(\dot\alpha_\wha\alpha_\whb-\dot\alpha_\whb\alpha_\wha)+\tfrac{1}{16}\beta^2\dot\gamma^{\wha\whb}\dot\gamma_{\wha\whb}\Big]
				\end{aligned}
			\end{equation}
			and
			\begin{equation}
				\begin{aligned}
					R_{\whmi\wha}\ &=\ \dot\alpha_\wha-\tfrac12\alpha^\whb\dot\gamma_{\wha\whb}+\tfrac14\alpha_\wha\dot\gamma_\whb{}^\whb-\tfrac12\tilde\nabla_\wha\dot\gamma_\whb{}^\whb+\tfrac12\tilde\nabla^\whb\dot\gamma_{\wha\whb}
					\\
					&\kern1cm+r\Big[\tfrac12\ddot\alpha_\wha+\tfrac12\alpha_\wha\ddot\gamma_\whb{}^\whb-\tfrac12\alpha_\whb\ddot\gamma_\wha{}^\whb-\tfrac12\dot\alpha_\whb\dot\gamma_\wha{}^\whb+\tfrac14\dot\alpha_\wha\dot\gamma_\whb{}^\whb-\tfrac14\alpha_\wha\dot\gamma^{\whb\whc}\dot\gamma_{\whb\whc}
					\\
					&\kern1cm+\tfrac12\alpha_\whb\Big(\dot\gamma^{\whb\whc}\dot\gamma_{\wha\whc}-\tfrac12\dot\gamma_\whc{}^\whc\dot\gamma_\wha{}^\whb\Big)\Big]
				\end{aligned}
			\end{equation}
			and
			\begin{equation}
				\begin{aligned}
					R_{\whpl\wha}\ &=\ r\Big(\beta\alpha_\wha-\tilde\nabla_\wha\beta-2\alpha^\whb\tilde\nabla_{[\wha}\alpha_{\whb]}+\tilde\nabla^\whb\tilde\nabla_{[\wha}\alpha_{\whb]}\Big)
					\\
					&\kern1cm+r^2\Big[-\tfrac12\beta\dot\alpha_\wha+2\dot\beta\alpha_\wha-\tfrac12\tilde\nabla_\wha\dot\beta-\tfrac32\alpha^\whb(\dot\alpha_\wha\alpha_\whb-\dot\alpha_\whb\alpha_\wha)+\alpha^\whb\tilde\nabla_\whb\dot\alpha_\wha
					\\
					&\kern1cm-\tfrac12\tilde\nabla_\wha(\dot\alpha^\whb\alpha_\whb)-\tfrac12\alpha_\wha\tilde\nabla^\whb\dot\alpha_\whb+\tfrac12\dot\alpha_\wha\tilde\nabla^\whb\alpha_\whb+\Big(-\tfrac34\beta\alpha_\whb+\tfrac12\tilde\nabla_\whb\beta\Big)\dot\gamma_\wha{}^\whb
					\\
					&\kern1cm+\Big(\tfrac34\beta\alpha_\wha-\tfrac12\tilde\nabla_\wha\beta\Big)\tfrac12\dot\gamma_\whb{}^\whb+\alpha^\whb\tilde\nabla_{[\whc}\alpha_{\whb]}\dot\gamma_\wha{}^\whc+\alpha^\whb\tilde\nabla_{[\wha}\alpha_{\whc]}\dot\gamma_\whb{}^\whc
					\\
					&\kern1cm-\tfrac12\alpha^\whb\tilde\nabla_{[\wha}\alpha_{\whb]}\dot\gamma_\whc{}^\whc-\tfrac14\beta \tilde\nabla_\wha\dot\gamma_\whb{}^\whb+\tfrac14\beta\tilde\nabla^\whb\dot\gamma_{\wha\whb}\Big]
					\\
					&\kern1cm+r^3\Big[-\tfrac14\beta\ddot\alpha_\wha+\tfrac12\ddot\beta\alpha_\wha+\tfrac12\alpha_\wha\alpha^\whb\ddot\alpha_\whb-\tfrac12\alpha^\whb\alpha_\whb\ddot\alpha_\wha-\tfrac12\dot\alpha_\wha\dot\alpha^\whb\alpha_\whb+\tfrac12\dot\alpha^\whb\dot\alpha_\whb\alpha_\wha
					\\
					&\kern1cm-\tfrac14\beta\alpha^\whb\ddot\gamma_{\wha\whb}+\tfrac14\beta\alpha_\wha\ddot\gamma_\whb{}^\whb+\Big(\tfrac14\beta\dot\alpha^\whb-\tfrac12\dot\beta\alpha^\whb\Big)\dot\gamma_{\wha\whb}+\Big(-\tfrac18\beta\dot\alpha_\wha +\tfrac14\dot\beta\alpha_\wha\Big)\dot\gamma_\whb{}^\whb
					\\
					&\kern1cm+\tfrac12\Big(\dot\alpha_\wha\alpha_\whb\alpha_\whc-\dot\alpha_\whc\alpha_\whb\alpha_\wha\Big)\dot\gamma^{\whb\whc}+\tfrac12\Big(\dot\alpha^\whc\alpha^\whb\alpha_\whb-\dot\alpha^\whb\alpha_\whb\alpha^\whc\Big)\dot\gamma_{\wha\whc}
					\\
					&\kern1cm+\tfrac14\Big(\dot\alpha^\whb\alpha_\whb\alpha_\wha-\dot\alpha_\wha\alpha^\whb\alpha_\whb\Big)\dot\gamma_\whc{}^\whc-\tfrac18\beta\alpha_\wha\dot\gamma^{\whb\whc}\dot\gamma_{\whb\whc}+\tfrac14\beta\alpha_\whb\Big(\dot\gamma^{\whb\whc}\dot\gamma_{\wha\whc}-\tfrac12\dot\gamma_\wha{}^\whb\dot\gamma_\whc{}^\whc\Big)\Big]
				\end{aligned}
			\end{equation}
			and
			\begin{equation}
				\begin{aligned}
					R_{\wha\whb}\ &=\ \tilde R_{\wha\whb}+\tilde\nabla_{(\wha}\alpha_{\whb)}-\tfrac12\alpha_\wha\alpha_\whb
					\\
					&\kern1cm+r\Big[\tilde\nabla_{(\wha}\dot\alpha_{\whb)}-\tfrac32\Big(\dot\alpha_\wha\alpha_\whb+\dot\alpha_\whb\alpha_\wha\Big)+\Big(\alpha^\whc\alpha_\whb-\tfrac12\tilde\nabla^\whc\alpha_\whb\Big)\dot\gamma_{\wha\whc}
					\\
					&\kern1cm+\Big(\alpha^\whc\alpha_\wha-\tfrac12\tilde\nabla^\whc\alpha_\wha\Big)\dot\gamma_{\whb\whc}+\Big(-\alpha^\whc\alpha_\whc-\beta+\tfrac12\tilde\nabla^\whc\alpha_\whc\Big)\dot\gamma_{\wha\whb}
					\\
					&\kern1cm+\tfrac12\Big(-\alpha_\wha\alpha_\whb+\tilde\nabla_{(\wha}\alpha_{\whb)}\Big)\dot\gamma_\whc{}^\whc-\alpha_{(\wha|}\tilde\nabla^\whc\dot\gamma_{\whc|\whb)}-\alpha^\whc\tilde\nabla_{(\wha|}\dot\gamma_{\whc|\whb)}
					\\
					&\kern1cm+\alpha^\whc\tilde\nabla_\whc\dot\gamma_{\wha\whb}+\alpha_{(\wha}\tilde\nabla_{\whb)}\dot\gamma_{\whc }{}^\whc
					\\
					&\kern1cm+r^2\Big[\alpha_{(\wha|}\alpha^\whc\ddot\gamma_{|\whb)\whc}-\ddot\alpha_{(\wha}\alpha_{\whb)}-\tfrac12\alpha_\wha\alpha_\whb\ddot\gamma_\whc{}^\whc-\tfrac12\Big(\alpha_\whc\alpha^\whc+\beta\Big)\ddot\gamma_{\wha\whb}-\tfrac12\dot\alpha_\wha\dot\alpha_\whb
					\\
					&\kern1cm-\Big(\tfrac12\dot\beta+\dot\alpha^\whc\alpha_\whc\Big)\dot\gamma_{\wha\whb}+\dot\alpha_{(\whc}\alpha_{\whb)}\dot\gamma_\wha{}^\whc+\dot\alpha_{(\whc}\alpha_{\wha)}\dot\gamma_\whb{}^\whc-\tfrac12\dot\alpha_{(\wha}\alpha_{\whb)}\dot\gamma_\whc{}^\whc
					\\
					&\kern1cm-\alpha_{(\wha|}\alpha_\whd\dot\gamma_{|\whb)\whc}\dot\gamma^{\whc\whd}+\tfrac12\alpha_{(\wha|}\alpha_\whc\dot\gamma_{|\whb)}{}^\whc\dot\gamma_\whd{}^\whd-\tfrac12\alpha_\whc\alpha_\whd\dot\gamma_\wha{}^\whc\dot\gamma_\whb{}^\whd+\tfrac12\Big(\alpha_\whd\alpha^\whd+\beta\Big)\dot\gamma_{\wha\whc}\dot\gamma_\whb{}^\whc
					\\
					&\kern1cm+\tfrac14\alpha_\wha\alpha_\whb\dot\gamma_{\whc\whd}\dot\gamma^{\whc\whd}+\tfrac12\alpha_\whc\alpha_\whd\dot\gamma^{\whc\whd}\dot\gamma_{\wha\whb}-\tfrac14\Big(\alpha_\whc\alpha^\whc+\beta\Big)\dot\gamma_{\wha\whb}\dot\gamma_\whd{}^\whd\Big]\,,
				\end{aligned}
			\end{equation}
		\end{subequations}
		where $\dot e_\wha{}^\whb$ is defined in~\eqref{eq:Defedot}, and $\tilde R_{\wha\whb}$ is the Ricci tensor with respect to $\tilde\nabla$.

		\section{Killing spinor equation}\label{App:KillingSpinorEqs}

		Recall the Killing spinor equation~\eqref{eq:KSE},
		\begin{equation}
			\nabla_\whA\epsilon+\Big(-\tfrac{1}{288}\Gamma_{\whA}{}^{\whB\whC\whD\whE}F_{\whB\whC\whD\whE}+\tfrac{1}{36}F_{\whA\whB\whC\whD}\Gamma^{\whB\whC\whD}\Big)\epsilon\ =\ 0~.
		\end{equation}
		This equation can be decomposed into the lightcone directions and those of the spatial cross-section. The $\whmi$ component is
		\begin{subequations}
			\begin{equation}\label{eq:P+KSE-}
				\partial_r\epsilon_\whpl\ =\ \tfrac14\dot e_\whb{}^\wha\Gamma_\wha{}^\whb\epsilon_\whpl-\tfrac{1}{24}Z^{\wha\whb\whc}\Gamma_{\wha\whb\whc}\epsilon_\whpl~,
			\end{equation}
			and
			\begin{equation}\label{eq:P-KSE-}
				\begin{aligned}
					\partial_r\epsilon_\whmi\ &=\ \Gamma_\whmi\Big[\tfrac14(\alpha^\wha+r\dot\alpha^\wha)\Gamma_\wha+\tfrac{1}{288}(X-r\alpha\wedge Z)^{\wha\whb\whc\whd}\Gamma_{\wha\whb\whc\whd}+\tfrac{1}{12}\Psi^{\wha\whb}\Gamma_{\wha\whb}\Big]\epsilon_\whpl
					\\
					&\kern1cm+\Big[\tfrac14\dot e_\whb{}^\wha\Gamma_\wha{}^\whb-\tfrac{1}{72}Z^{\wha\whb\whc}\Gamma_{\wha\whb\whc}\Big]\epsilon_\whmi~.
				\end{aligned}
			\end{equation}
			The $\whpl$ component is
			\begin{equation}
				\begin{aligned}
					\tfrac12r^2\beta\partial_r\epsilon_\whpl+\partial_u\epsilon_\whpl\ &=\ \Big[-\tfrac12\Big(r\beta+\tfrac12r^2\dot\beta\Big)+\tfrac14\Big(r\tilde\nabla^\wha\alpha_\whb+r^2 \dot\alpha^\wha\alpha_\whb+\tfrac12r^2\beta\dot e_\whb{}^\wha\Big)\Gamma_\wha{}^\whb
					\\
					&\kern1cm-\tfrac{1}{72}r(W-\alpha\wedge\Psi+\tfrac12r\beta Z)_{\whb_1\whb_2\whb_3}\Gamma^{\whb_1\whb_2\whb_3}\Big]\epsilon_\whpl
					\\
					&\kern1cm+\Gamma_\whpl\Big[\tfrac14(\alpha_\wha+r\dot\alpha_\wha)\Gamma^\wha+\tfrac{1}{288}(X-r\alpha\wedge Z)^{\wha\whb\whc\whd}\Gamma_{\wha\whb\whc\whd}
					\\
					&\kern1cm-\tfrac{1}{12}\Psi^{\wha\whb}\Gamma_{\wha\whb}\Big]\epsilon_-
				\end{aligned}
			\end{equation}
			and
			\begin{equation}
				\begin{aligned}
					\tfrac12r^2\beta\partial_r\epsilon_\whmi +\partial_u \epsilon_\whmi\ &=\ \tfrac12\Big(-\tfrac12r^2\beta\alpha^\wha+\tfrac12r^2\tilde\nabla^\wha\beta+\tfrac12r^3\beta\dot\alpha^\wha-\tfrac12r^3\dot\beta\alpha^\wha\Big)\Gamma_\whmi\Gamma_\wha\epsilon_\whpl
					\\
					&\kern1cm+\Big[\tfrac12\Big(r\beta+\tfrac12r^2\dot\beta\Big)+\tfrac14\Big(r\tilde\nabla^\wha\alpha_\whb+r^2 \dot\alpha^\wha\alpha_\whb+\tfrac12r^2\beta\dot e_\whb{}^\wha\Big)\Gamma_\wha{}^\whb
					\\
					&\kern1cm-\tfrac{1}{24}r(W-\alpha\wedge\Psi+\tfrac12r\beta Z)^{\wha\whb\whc}\Gamma_{\wha\whb\whc}\Big]\epsilon_\whmi~.
				\end{aligned}
			\end{equation}
			The $\wha$ components are
			\begin{equation}
				\begin{aligned}
					0\ &=\ -r\alpha_\wha\partial_r\epsilon_\whmi+\tilde\nabla_\wha\epsilon_\whmi
					\\
					&\kern1cm+\Gamma_\whmi\Big[-\tfrac14\Big(r(\tilde\rmd\alpha)_{\wha\whb}+2r^2\dot\alpha_{[\wha}\alpha_{\whb]}+r^2\beta \dot e_{(\whb\wha)}\Big)\Gamma^\whb
					\\
					&\kern1cm+\tfrac{1}{72}r(W-\alpha\wedge\Psi+\tfrac12r\beta Z)_{\wha\whb\whc}\Gamma_\wha{}^{\wha\whb\whc}-\tfrac{1}{12}r(W-\alpha\wedge\Psi+\tfrac12r\beta Z)_{\wha\whb\whc}\Gamma^{\whb\whc}\Big]\epsilon_\whpl
					\\
					&\kern1cm+\Big[\tfrac14(\alpha_\wha+r\dot\alpha_\wha)-\tfrac14r(2\alpha_{[\wha}\dot e_{\whb]\whc}+\alpha_\whc\dot e_{\whb\wha})\Gamma^{\whb\whc}-\tfrac{1}{24}\Psi_{\whb\whc}\Gamma_\wha{}^{\whb\whc}+\tfrac16\Psi_{\wha\whb}\Gamma^\whb
					\\
					&\kern1cm-\tfrac{1}{288}(X-r\alpha\wedge Z)_{\whb\whc\whd\whe}\Gamma_\wha{}^{\whb\whc\whd\whe}+\tfrac{1}{36}(X-r\alpha\wedge Z)_{\wha\whb\whc\whd}\Gamma^{\whb\whc\whd}\Big]\epsilon_\whmi
				\end{aligned}
			\end{equation}
			and
			\begin{equation}
				\begin{aligned}
					0\ &=\ -r\alpha_\wha\partial_r\epsilon_\whpl+\tilde\nabla_\wha\epsilon_\whpl
					\\
					&\kern1cm+\Big[-\tfrac14(\alpha_\wha+r\dot\alpha_\wha)-\tfrac14r(2\alpha_{[\wha}\dot e_{\whb]\whc}+\alpha_\whc\dot e_{\whb\wha})\Gamma^{\whb\whc}+\tfrac{1}{24}\Psi_{\whb\whc}\Gamma_\wha{}^{\whb\whc}-\tfrac16\Psi_{\wha\whb}\Gamma^\whb
					\\
					&\kern1cm-\tfrac{1}{288}(X-r\alpha\wedge Z)_{\whb\whc\whd\whe}\Gamma_\wha{}^{\whb\whc\whd\whe}+\tfrac{1}{36}(X-r\alpha\wedge Z)_{\wha\whb\whc\whd}\Gamma^{\whb\whc\whd}\Big]\epsilon_\whpl
					\\
					&\kern1cm+\Gamma_\whpl\Big[-\tfrac12\dot e_{(\whb\wha)}\Gamma^\whb+\tfrac{1}{72}Z_{\whb\whc\whd}\Gamma_\wha{}^{\whb\whc\whd}-\tfrac{1}{12}Z_{\wha\whb\whc}\Gamma^{\whb\whc}\Big]\epsilon_\whmi~,
				\end{aligned}
			\end{equation}
		\end{subequations}
		where $\tilde\nabla$ is the Levi-Civita connection with respect to the metric $\gamma$ on the spatial cross-section $S$ and $\dot e_\wha{}^\whb$ is defined in~\eqref{eq:Defedot}.
		
		For convenience, we define 
		\begin{equation}\label{eq:DefTheta}
			\begin{aligned}
				\Theta_{\widehat\pm}\ &\coloneqq\ \tfrac14(\alpha^\wha+r\dot\alpha^\wha)\Gamma_\wha+\tfrac{1}{288}(X-r\alpha\wedge Z)^{\wha\whb\whc\whd}\Gamma_{\wha\whb\whc\whd}\pm\tfrac{1}{12}\Psi^{\wha\whb}\Gamma_{\wha\whb}~,
				\\
				\bar\Theta_{\widehat\pm}\ &\coloneqq\ -\tfrac14(\alpha^\wha+r\dot\alpha^\wha)\Gamma_\wha+\tfrac{1}{288}(X-r\alpha\wedge Z)^{\wha\whb\whc\whd}\Gamma_{\wha\whb\whc\whd}\pm\tfrac{1}{12}\Psi^{\wha\whb}\Gamma_{\wha\whb}~.
			\end{aligned}
		\end{equation}
		
		Using the $\whmi$ component~\eqref{eq:P-KSE-} and~\eqref{eq:P+KSE-}, we can get rid of the $r$-derivatives of the spinor. The Killing spinor equation simplifies to
		\begin{subequations}\label{eq:AllKSE}
			\begin{equation}\label{eq:KSEr+}
				\partial_r\epsilon_\whpl\ =\ \tfrac14\dot e_\whb{}^\wha\Gamma_\wha{}^\whb\epsilon_\whpl-\tfrac{1}{24}Z^{\wha\whb\whc}\Gamma_{\wha\whb\whc}\epsilon_\whpl
			\end{equation}
			and
			\begin{equation}\label{eq:KSEr-}
				\begin{aligned}
					\partial_r\epsilon_\whmi\ &=\ \Gamma_\whmi\Theta_\whpl\epsilon_\whpl+\Big[\tfrac14\dot e_\whb{}^\wha\Gamma_\wha{}^\whb-\tfrac{1}{72}Z^{\wha\whb\whc}\Gamma_{\wha\whb\whc}\Big]\epsilon_\whmi
				\end{aligned}
			\end{equation}
			and
			\begin{equation}\label{eq:KSEu+}
				\begin{aligned}
					\partial_u\epsilon_\whpl\ &=\ \Big[-\tfrac12\Big(r\beta+\tfrac12r^2\dot\beta\Big)+\tfrac14\Big(r\tilde\nabla^\wha\alpha_\whb+r^2\dot\alpha^\wha\alpha_\whb\Big)\Gamma_\wha{}^\whb
					\\
					&\kern1cm-\tfrac{1}{72}r(W-\alpha\wedge\Psi-r\beta Z)_{\wha\whb\whc}\Gamma^{\wha\whb\whc}\Big]\epsilon_\whpl+\Gamma_\whpl\Theta_\whmi\epsilon_\whmi
				\end{aligned}
			\end{equation}
			and
			\begin{equation}
				\begin{aligned}
					\partial_u\epsilon_\whmi\ &=\ \Gamma_\whmi\Big[\Big(-\tfrac14r^2\beta\alpha^\wha+\tfrac14r^2\tilde\nabla^\wha\beta+
					\tfrac14r^3\dot\beta\alpha^\wha\Big)\Gamma_\wha-\tfrac12r^2\beta\Theta_\whpl\Big]\epsilon_\whpl
					\\
					&\kern1cm+\Big[\tfrac12\Big(r\beta+\tfrac12r^2\dot\beta\Big)+\tfrac14\Big(r\tilde\nabla^\wha\alpha_\whb+r^2 \dot\alpha^\wha\alpha_\whb\Big)\Gamma_\wha{}^\whb
					\\
					&\kern1cm-\tfrac{1}{24}r(W-\alpha\wedge\Psi+\tfrac13r\beta Z)^{\wha\whb\whc}\Gamma_{\wha\whb\whc}\Big]\epsilon_\whmi
				\end{aligned}
			\end{equation}
			and
			\begin{equation}\label{eq:KSEa-}
				\begin{aligned}
					0\ &=\  \tilde\nabla_\wha\epsilon_\whmi
					\\
					&\kern1cm+\Gamma_\whmi\Big[-\tfrac14\Big(r(\tilde\rmd\alpha)_{\wha\whb}+2r^2\dot\alpha_{[\wha}\alpha_{\whb]}+\tfrac12r^2\beta\dot\gamma_{\wha\whb}\Big)\Gamma^\whb-r\alpha_a\Theta_\whpl
					\\
					&\kern1cm+\tfrac{1}{72}r(W-\alpha\wedge\Psi+\tfrac12r\beta Z)_{\whb\whc\whd}\Gamma_\wha{}^{\whb\whc\whd}-\tfrac{1}{12}r(W-\alpha\wedge\Psi+\tfrac12r\beta Z)_{\wha\whb\whc}\Gamma^{\whb\whc}\Big]\epsilon_\whpl
					\\
					&\kern1cm+\Big[\tfrac14(\alpha_\wha+r\dot\alpha_\wha)+\tfrac14r\alpha_\whb\dot\gamma_{\wha\whc}\Gamma^{\whb\whc}-\tfrac{1}{24}\Psi_{\whb\whc}\Gamma_\wha{}^{\whb\whc}+\tfrac16\Psi_{\wha\whb}\Gamma^\whb
					\\
					&\kern1cm-\tfrac{1}{288}(X-r\alpha\wedge Z)_{\whb\whc\whd\whe}\Gamma_\wha{}^{\whb\whc\whd\whe}+\tfrac{1}{36}(X-r\alpha\wedge Z)_{\wha\whb\whc\whd}\Gamma^{\whb\whc\whd}
					\\
					&\kern1cm+\tfrac{1}{72}r\alpha_\wha Z^{\whb\whc\whd}\Gamma_{\whb\whc\whd}\Big]\epsilon_\whmi
				\end{aligned}
			\end{equation}
			and
			\begin{equation}\label{eq:KSEa+}
				\begin{aligned}
					0\ &=\ \tilde\nabla_\wha\epsilon_\whpl
					\\
					&\kern1cm+\Big[-\tfrac14\Big(\alpha_\wha+r\dot\alpha_\wha\Big)+\tfrac14r\alpha_\whb\dot\gamma_{\wha\whc}\Gamma^{\whb\whc}+\tfrac{1}{24}\Psi_{\whb\whc}\Gamma_\wha{}^{\whb\whc}-\tfrac16\Psi_{\wha\whb}\Gamma^\whb
					\\
					&\kern1cm-\tfrac{1}{288}(X-r\alpha\wedge Z)_{\whb\whc\whd\whe}\Gamma_\wha{}^{\whb\whc\whd\whe}+\tfrac{1}{36}(X-r\alpha\wedge Z)_{\wha\whb\whc\whd}\Gamma^{\whb\whc\whd}
					\\
					&\kern1cm+\tfrac{1}{24}r\alpha_\wha Z^{\whb\whc\whd}\Gamma_{\whb\whc\whd}\Big]\epsilon_\whpl
					\\
					&\kern1cm+\Gamma_\whpl\Big(-\tfrac14\dot\gamma_{\wha\whb}\Gamma^\whb+\tfrac{1}{72}Z_{\whb\whc\whd}\Gamma_\wha{}^{\whb\whc\whd}-\tfrac{1}{12}Z_{\wha\whb\whc}\Gamma^{\whb\whc}\Big)\epsilon_\whmi~.
				\end{aligned}
			\end{equation}
		\end{subequations}

		\section{Spinors from forms}\label{App:SpinorForm}

		Following~\cite{Gran:2005wu}, we can view spinors as differential forms. In particular, let $U\coloneqq\IR^5$ with an orthonormal basis $\bfe_1,\ldots,\bfe_5$ and Euclidean metric. On the complexification $U_\IC\coloneqq U\otimes\IC$, we can introduce the standard Hermitian inner product
		\begin{equation}
			\inner{z}{w}\ \coloneqq\ \sum_{m=1}^{5}(z^m)^*w^m~,
		\end{equation}
		for all $z=z^m\bfe_m\in U_\IC$ and $w=w^m\bfe_m\in U_\IC$ and $(z^m)^*$ is the complex conjugate of $z^m$. This inner product can be extended to the exterior algebra $\Lambda^\bullet U_\IC$, and we shall use the same notation $\inner{-}{-}$ for this extension. Spinors are then elements of $\Lambda^\bullet U_\IC$. We now introduce two representations of the gamma matrices, called the timelike representation and the null representation, which get their respective names from spinors whose bilinear one-forms are timelike and null, respectively.

		\paragraph{Timelike representation.}
		In the timelike representation, the gamma matrices act on spinors $\epsilon\in\Lambda^\bullet U_\IC$ as
		\begin{equation}
			\Gamma_\ulm\epsilon\ \coloneqq\ \bfe_m\wedge\epsilon+\bfe_m\intprod\epsilon
			\eand
			\Gamma_{\underline{5+m}}\epsilon\ \coloneqq\ \rmi\bfe_m\wedge\epsilon-\rmi\bfe_m\intprod\epsilon
		\end{equation}
		for all $m=1,\ldots,5$ and where $\bfe_m\intprod$ is the adjoint of $\bfe_m\wedge$ with respect to $\inner{-}{-}$ and `i' the imaginary unit. We set $\Gamma_{\underline{0}}\coloneqq\Gamma_{\underline{1}}\cdots\Gamma_{\underline{10}}$. Then, $\{\Gamma_\ulA,\Gamma_\ulB\}=2g_{\ulA\ulB}$ with $g_{\ulA\ulB}=\eta_{\ulA\ulB}$ and $\diag(\eta_{\ulA\ulB})=(-1,1,\ldots,1)$.

		With $B\coloneqq\Gamma_{\underline{6}}\cdots\Gamma_{\underline{10}}$, the Majorana inner product is
		\begin{equation}\label{eq:timelikeOneFormBilinear}
			B(\epsilon,\eta)\ \coloneqq\ \inner{B\epsilon^*}{\eta}
		\end{equation}
		for all $\epsilon,\eta\in\Lambda^\bullet U_\IC$.

		Finally, the Majorana condition is 
		\begin{equation}
			\epsilon^*\ =\ \Gamma_{\underline{0}}B\epsilon~.
		\end{equation}

		\paragraph{Null representation.}
		In the null representation, the gamma matrices act on spinors $\epsilon\in\Lambda^\bullet U_\IC$ as
		\begin{equation}
			\begin{gathered}
				\Gamma_\whpl\epsilon\ \coloneqq\ \sqrt{2}\bfe_5\intprod\epsilon~,
				\quad
				\Gamma_\whmi\epsilon\ \coloneqq\ -\sqrt{2}\bfe_5\wedge\epsilon~,
				\\
				\Gamma_\whm\epsilon\ \coloneqq\ \bfe_m\wedge\epsilon+\bfe_m\intprod\epsilon~,
				\quad
				1\ \leq\ m\ \leq\ 4~,
				\\
				\Gamma_{\widehat{5+m}}\epsilon\ \coloneqq\ \rmi\bfe_m\wedge\epsilon-\rmi\bfe_m\intprod\epsilon~,
				\quad
				1\ \leq\ m\ \leq\ 4~,
			\end{gathered}
		\end{equation}
		where $\bfe_m\intprod$ is the adjoint of $\bfe_m\wedge$ with respect to $\inner{-}{-}$ and `i' the imaginary unit. Then, $\{\Gamma_\whA,\Gamma_\whA\}=2g_{\whA\whB}$ with $g_{\whA\whB}$ given in~\eqref{eq:MetricOrthonormal}.

		With $B\coloneqq\frac{1}{\sqrt{2}}(\Gamma_\whpl-\Gamma_\whmi)\Gamma_{\widehat6}\Gamma_{\widehat7}\Gamma_{\widehat8}\Gamma_{\widehat9}$, the Majorana inner product is
		\begin{equation}
			B(\epsilon,\eta)\ \coloneqq\ \inner{B\epsilon^*}{\eta}
		\end{equation}
		for all $\epsilon,\eta\in\Lambda^\bullet U_\IC$.

		Finally, the Majorana condition is
		\begin{equation}
			\epsilon^*\ =\ -\tfrac{1}{\sqrt{2}}(\Gamma_\whpl-\Gamma_\whmi)B\epsilon~.
		\end{equation}

	\end{body}


\begin{thebibliography}{10}

\bibitem{Gauntlett:2002nw}
J.~P.~Gauntlett, J.~B.~Gutowski, C.~M.~Hull, S.~Pakis, and H.~S.~Reall,
{\em {All supersymmetric solutions of minimal supergravity in five
  dimensions},}
\href{https://dx.doi.org/10.1088/0264-9381/20/21/005}{Class. Quant. Grav. {\bf
  20}  (2003) 4587} [{\tt
  \href{https://www.arxiv.org/abs/hep-th/0209114}{hep-th/0209114}}].

\bibitem{Gauntlett:2003fk}
J.~P.~Gauntlett and J.~B.~Gutowski,
{\em {All supersymmetric solutions of minimal gauged supergravity in
  five-dimensions},}
\href{https://dx.doi.org/10.1103/PhysRevD.70.089901}{Phys. Rev. D {\bf 68}
  (2003) 105009} [{\tt
  \href{https://www.arxiv.org/abs/hep-th/0304064}{hep-th/0304064}}].

\bibitem{Lucietti:2023mvj}
J.~Lucietti, P.~Ntokos, and S.~G.~Ovchinnikov,
{\em {All separable supersymmetric AdS$_{5}$ black holes},}
\href{https://dx.doi.org/10.1007/JHEP05(2024)062}{JHEP {\bf 05}  (2024) 062}
  [{\tt \href{https://www.arxiv.org/abs/2311.06124}{2311.06124 [hep-th]}}].

\bibitem{Dias:2023rde}
O.~J.~C.~Dias, G.~W.~Gibbons, J.~E.~Santos, and B.~Way,
{\em {Static black binaries in de Sitter space},}
\href{https://dx.doi.org/10.1103/PhysRevLett.131.131401}{Phys. Rev. Lett. {\bf
  131}  (2023) 131401} [{\tt
  \href{https://www.arxiv.org/abs/2303.07361}{2303.07361 [gr-qc]}}].

\bibitem{Hawking:1971vc}
S.~W.~Hawking,
{\em {Black holes in general relativity},}
\href{https://dx.doi.org/10.1007/BF01877517}{Commun. Math. Phys. {\bf 25}
  (1972) 152}.

\bibitem{Hawking:1973uf}
S.~W.~Hawking and G.~F.~R.~Ellis,
{\em {The large scale structure of space-time},} Cambridge University Press,
  2023
[\href{https://dx.doi.org/10.1017/9781009253161}{doi}].

\bibitem{Hollands:2008wn}
S.~Hollands and A.~Ishibashi,
{\em {On the `stationary implies axisymmetric' theorem for extremal black holes
  in higher dimensions},}
\href{https://dx.doi.org/10.1007/s00220-009-0841-1}{Commun. Math. Phys. {\bf
  291}  (2009) 403} [{\tt \href{https://www.arxiv.org/abs/0809.2659}{0809.2659
  [gr-qc]}}].

\bibitem{Dunajski:2023xrd}
M.~Dunajski and J.~Lucietti,
{\em {Intrinsic rigidity of extremal horizons},}
\href{https://dx.doi.org/10.4310/jdg/1766434274}{J. Diff. Geom. {\bf 132}
  (2026) 179} [{\tt \href{https://www.arxiv.org/abs/2306.17512}{2306.17512
  [gr-qc]}}].

\bibitem{Colling:2024usk}
A.~Colling, M.~Dunajski, H.~Kunduri, and J.~Lucietti,
{\em {New quasi-Einstein metrics on a two-sphere},}
\href{https://dx.doi.org/10.1007/s12220-026-02459-0}{J. Geom. Anal. {\bf 36}
  (2026) 226} [{\tt \href{https://www.arxiv.org/abs/2403.04117}{2403.04117
  [math.DG]}}].

\bibitem{Colling:2025dub}
A.~Colling,
{\em {Symmetries of extremal horizons},}
{\tt \href{https://www.arxiv.org/abs/2512.10200}{2512.10200 [gr-qc]}}.

\bibitem{Chrusciel:2017vie}
P.~T.~Chru{\'s}ciel, S.~J.~Szybka, and P.~Tod,
{\em {Towards a classification of vacuum near-horizons geometries},}
\href{https://dx.doi.org/10.1088/1361-6382/aa90e7}{Class. Quant. Grav. {\bf 35}
   (2018) 015002} [{\tt \href{https://www.arxiv.org/abs/1707.01118}{1707.01118
  [gr-qc]}}].

\bibitem{Kunduri:2013gce}
H.~K.~Kunduri and J.~Lucietti,
{\em {Classification of near-horizon geometries of extremal black holes},}
\href{https://dx.doi.org/10.12942/lrr-2013-8}{Living Rev. Rel. {\bf 16}
  (2013)~8} [{\tt \href{https://www.arxiv.org/abs/1306.2517}{1306.2517
  [hep-th]}}].

\bibitem{Gutowski:2013kma}
J.~Gutowski and G.~Papadopoulos,
{\em {Index theory and dynamical symmetry enhancement of M-horizons},}
\href{https://dx.doi.org/10.1007/JHEP05(2013)088}{JHEP {\bf 05}  (2013) 088}
  [{\tt \href{https://www.arxiv.org/abs/1303.0869}{1303.0869 [hep-th]}}].

\bibitem{Moncrief:1983xua}
V.~Moncrief and J.~Isenberg,
{\em {Symmetries of cosmological Cauchy horizons},}
\href{https://dx.doi.org/10.1007/BF01214662}{Commun. Math. Phys. {\bf 89}
  (1983) 387}.

\bibitem{Gillard:2004xq}
J.~Gillard, U.~Gran, and G.~Papadopoulos,
{\em {The spinorial geometry of supersymmetric backgrounds},}
\href{https://dx.doi.org/10.1088/0264-9381/22/6/009}{Class. Quant. Grav. {\bf
  22}  (2005) 1033} [{\tt
  \href{https://www.arxiv.org/abs/hep-th/0410155}{hep-th/0410155}}].

\bibitem{Gauntlett:2003wb}
J.~P.~Gauntlett, J.~B.~Gutowski, and S.~Pakis,
{\em {The deometry of $D=11$ null Killing spinors},}
\href{https://dx.doi.org/10.1088/1126-6708/2003/12/049}{JHEP {\bf 12}  (2003)
  049} [{\tt \href{https://www.arxiv.org/abs/hep-th/0311112}{hep-th/0311112}}].

\bibitem{Figueroa-OFarrill:2002ecq}
J.~M.~Figueroa-O'Farrill and G.~Papadopoulos,
{\em {Maximally supersymmetric solutions of ten-dimensional and
  eleven-dimensional supergravities},}
\href{https://dx.doi.org/10.1088/1126-6708/2003/03/048}{JHEP {\bf 03}  (2003)
  048} [{\tt \href{https://www.arxiv.org/abs/hep-th/0211089}{hep-th/0211089}}].

\bibitem{Gran:2006cn}
U.~Gran, J.~Gutowski, G.~Papadopoulos, and D.~Roest,
{\em $N=31$, $D=11$,}
\href{https://dx.doi.org/10.1088/1126-6708/2007/02/043}{JHEP {\bf 02}  (2007)
  043} [{\tt \href{https://www.arxiv.org/abs/hep-th/0610331}{hep-th/0610331}}].

\bibitem{Gran:2010tj}
U.~Gran, J.~Gutowski, and G.~Papadopoulos,
{\em {M-theory backgrounds with 30 Killing spinors are maximally
  supersymmetric},}
\href{https://dx.doi.org/10.1007/JHEP03(2010)112}{JHEP {\bf 03}  (2010) 112}
  [{\tt \href{https://www.arxiv.org/abs/1001.1103}{1001.1103 [hep-th]}}].

\bibitem{DiBella:2026vyi}
E.~Di~Bella, W.~A.~de~Graaf, and A.~Santi,
{\em {Some rigidity results for supergravity backgrounds in 11 dimensions},}
{\tt \href{https://www.arxiv.org/abs/2603.19923}{2603.19923 [hep-th]}}.

\bibitem{Figueroa-OFarrill:2012kws}
J.~Figueroa-O'Farrill and N.~Hustler,
{\em {The homogeneity theorem for supergravity backgrounds},}
\href{https://dx.doi.org/10.1007/JHEP10(2012)014}{JHEP {\bf 10}  (2012) 014}
  [{\tt \href{https://www.arxiv.org/abs/1208.0553}{1208.0553 [hep-th]}}].

\bibitem{Cremmer:1978km}
E.~Cremmer, B.~Julia, and J.~Scherk,
{\em Supergravity theory in 11 dimensions,}
\href{https://dx.doi.org/10.1016/0370-2693(78)90894-8}{Phys. Lett. B {\bf 76}
  (1978) 409}.

\bibitem{Hubeny:2002pj}
V.~E.~Hubeny and M.~Rangamani,
{\em {No horizons in pp waves},}
\href{https://dx.doi.org/10.1088/1126-6708/2002/11/021}{JHEP {\bf 11}  (2002)
  021} [{\tt \href{https://www.arxiv.org/abs/hep-th/0210234}{hep-th/0210234}}].

\bibitem{Hubeny:2003ug}
V.~E.~Hubeny and M.~Rangamani,
{\em {Horizons and plane waves: a review},}
\href{https://dx.doi.org/10.1142/S0217732303012428}{Mod. Phys. Lett. A {\bf 18}
   (2003) 2699} [{\tt
  \href{https://www.arxiv.org/abs/hep-th/0311053}{hep-th/0311053}}].

\bibitem{Gutowski:2025lzi}
J.~Gutowski, C.~Saelim, and M.~Wolf,
{\em {Extremal black holes from homotopy algebras},}
\href{https://dx.doi.org/10.1088/1361-6382/ae3046}{Class. Quant. Grav. {\bf 43}
   (2026) 025002} [{\tt \href{https://www.arxiv.org/abs/2508.08886}{2508.08886
  [hep-th]}}].

\bibitem{Fontanella:2016lzo}
A.~Fontanella and J.~B.~Gutowski,
{\em {Moduli spaces of transverse deformations of near-horizon geometries},}
\href{https://dx.doi.org/10.1088/1751-8121/aa6cbf}{J. Phys. A {\bf 50}  (2017)
  215202} [{\tt \href{https://www.arxiv.org/abs/1610.09949}{1610.09949
  [hep-th]}}].

\bibitem{Gran:2005wu}
U.~Gran, G.~Papadopoulos, and D.~Roest,
{\em {Systematics of M-theory spinorial geometry},}
\href{https://dx.doi.org/10.1088/0264-9381/22/13/013}{Class. Quant. Grav. {\bf
  22}  (2005) 2701} [{\tt
  \href{https://www.arxiv.org/abs/hep-th/0503046}{hep-th/0503046}}].

\bibitem{PUCCI20041}
P.~Pucci and J.~Serrin,
{\em The strong maximum principle revisited,}
\href{https://dx.doi.org/10.1016/j.jde.2003.05.001}{J. Differential Equations
  {\bf 196}  (2004)~1}.

\bibitem{Farotti:2021otm}
D.~Farotti and J.~Gutowski,
{\em {$N=4$ near-horizon geometries in $D=11$ supergravity},}
\href{https://dx.doi.org/10.1007/JHEP07(2021)155}{JHEP {\bf 07}  (2021) 155}
  [{\tt \href{https://www.arxiv.org/abs/2104.05478}{2104.05478 [hep-th]}}].

\bibitem{besse2007einstein}
A.~L.~Besse,
{\em Einstein manifolds,} Springer, 2007
[\href{https://dx.doi.org/10.1007/978-3-540-74311-8}{doi}].

\bibitem{Blau:2002mw}
M.~Blau, J.~M.~Figueroa-O'Farrill, and G.~Papadopoulos,
{\em {Penrose limits, supergravity and brane dynamics},}
\href{https://dx.doi.org/10.1088/0264-9381/19/18/310}{Class. Quant. Grav. {\bf
  19}  (2002) 4753} [{\tt
  \href{https://www.arxiv.org/abs/hep-th/0202111}{hep-th/0202111}}].

\end{thebibliography}
\end{document}